\documentclass[aps,prb,superscriptaddress,twocolumn]{revtex4}

\usepackage{dcolumn}
\usepackage{bm}
\usepackage{amsfonts}
\usepackage{graphicx,graphics}
\usepackage{amsmath,amssymb}
\usepackage{rotate}
\usepackage{subfigure}
\usepackage{epsfig}
\usepackage{tabularx,hhline}
\usepackage{theorem}
\usepackage{wasysym}
\usepackage{color}
\usepackage{times}

\begin{document}

\title{Itinerant electrons in the Coulomb phase}

\author{L.~D.~C.~Jaubert}
\affiliation{Max-Planck-Institut f\"ur Physik komplexer Systeme, 01187 Dresden, Germany.} 

\author{Swann Piatecki}
\affiliation{Max-Planck-Institut f\"ur Physik komplexer Systeme, 01187 Dresden, Germany.} 
\affiliation{Laboratoire de Physique Statistique, \'Ecole Normale
Sup\'erieure, UPMC, Universit\'e Paris Diderot, CNRS, 24 rue Lhomond, 75005 Paris, France.}

\author{Masudul Haque}
\affiliation{Max-Planck-Institut f\"ur Physik komplexer Systeme, 01187 Dresden, Germany.} 

\author{R.~Moessner}
\affiliation{Max-Planck-Institut f\"ur Physik komplexer Systeme, 01187 Dresden, Germany.}

\date{\today}

\begin{abstract}
We study the interplay between magnetic frustration and itinerant electrons. For example, how does the coupling to mobile charges modify the properties of a spin liquid, and does the underlying frustration favor insulating or conducting states?
Supported by Monte Carlo simulations, our goal is in particular to provide an analytical picture of the mechanisms involved.
The models under considerations exhibit Coulomb phases in two and three dimensions, where the itinerant electrons are coupled to the localized spins via double exchange interactions. Because of the Hund coupling, magnetic loops naturally emerge from the Coulomb phase and serve as conducting channels for the mobile electrons, leading to doping-dependent rearrangements of the loop ensemble in order to minimize the electronic kinetic energy.
At low electron density $\rho$, the double exchange coupling mainly tends to segment the very long loops winding around the system into smaller ones while it gradually lifts the extensive degeneracy of the Coulomb phase with increasing $\rho$. For higher doping, the results are strongly lattice dependent, displaying loop crystals with a given loop length for some specific values of $\rho$, which can melt into another loop crystal by varying $\rho$.
Finally, we contrast this to the qualitatively different behavior of analogous models on kagome or triangular lattices.
\end{abstract}


\maketitle

\section{Introduction}

The combination of magnetism and itinerant
electrons is a multi-faceted field in the physics of correlated
electrons, where our understanding is still remarkably patchy: even in the case of a square lattice Hubbard model, we lack consensus 
on a detailed phase diagram in the doping-temperature plane. 

Besides the cuprate superconductors, there are plenty of other
settings in which interesting questions arise, not least popularized
of late by questions raised by the pnictide superconductors, where
magnetic frustration and accidental degeneracies have started to be considered.

More broadly, there has been increased interest in the interaction of
frustrated magnetism with itinerant electrons~\cite{Ohgushi00a,Onoda03b,Ikoma03a,Haerter05a,Shimomura05a,Martin08a,Kalitsov09a,Motome10b,Kumar10a,Udagawa10a,Tang11a}.
Here, we take up the spirit of this thread of work and study itinerant
electrons on a highly frustrated lattice, the pyrochlore lattice. We
consider both three and two dimensions, the latter case also being
known as the square lattice with crossings, planar pyrochlore, or
checkerboard lattice.

We start with an exotic frustrated phase of a magnetic insulator, the Coulomb phase, which has been extensively studied recently~\cite{Huse03a,Chen09a,Morris09a,Fennell09a,Henley10a,Powell11a}. This phase has a number of unusual properties, including algebraic spin correlations and the emergence of extended one-dimensional degrees of freedom~\cite{Villain79a,Banks11a,Jaubert11a}, the nature of which is an independently interesting problem~\cite{Nahum11a}.

Especially the latter will play an important role in the following
analysis, given the ability of electrons to provide evidence for
non-local structures through properties related to transport
phenomena. Indeed, it is the marriage of the local constraints imposed
by frustration with the `non-local' physics describing mobile particles
which makes up for much of the interest in this field.

The Ising-double exchange model which we study here has many parameters: electron density,
$\rho$, temperature $T$, Ising anisotropy, Hund's coupling, $J_H$,
magnetic exchange, $J$, and electron hopping integral, $t$. A full study of
general parameter choices is well-nigh impossible in any detail analytically.

Our approach to the problem considers a regime where the effects of frustration are particularly strong but where considerable progress towards a detailed description is
nonetheless possible by analytical (or simple numerical) means.  That a (non-trivial) regime where this is possible exists at all is a priori not obvious, and
we find that we need to restrict a number of parameters to limiting
values: we study the limit where the magnetic energy scales are much
larger than the hopping integral $t$, so that the resulting problem is
one of electrons hopping on a classical background spin
configuration. We do not need to restrict the electron density to be
small, although for that case, we have the most detailed set of
results.

We find that the low-density behaviour can be mapped onto a study of a
classical loop model with non-trivial weights arising from the
addition of electrons Hund's coupled to the spins in the Coulomb
phase.  
This results in phenomena such as a transition from a (in
two dimensions, critical) percolation situation to one in which the loops acquire an
exponential length distribution, thereby removing all
conducting paths across the sample.
As the doping is increased further, we find a sequence of density-dependent
preferred loop lengths, which lead to a tendency to form loop crystals which
may, however, be frustrated by the lattice geometry.

The organization of this article, and our main results, are summarized below.

\subsection{Summary and Overview}

We restrict ourselves to the limit of large exchange coupling interactions so that
the ice-rules (see figure~\ref{fig:latt}) themselves are never compromised.  We make extensive use of the loop picture encoding the ice rules. \cite{Jaubert11a} The loops serve as 1D
channels for the electrons.  The problem is thus transformed to entropy and
energy considerations of possible loop coverings, with loops supporting
varying numbers of electron.  This ``loop framework'' for describing
conduction electrons is introduced in Section \ref{sec:loop_framework}.  

Ref.~\onlinecite{Jaubert11a} has described in some detail the loop distributions
for both 2D and 3D in the absence of electrons.  
Once electrons are added, within each loop electrons can occupy states
whose energy is given by a 1D dispersion.  The dispersion minimum is independent of loop
length, so the first electron in a loop has the same energy in all loops.  As
a result, there is a low doping regime where it is possible to fit at most one
electron per loop.  All such configurations have the same energy.
This is the entropic regime, because entropic arguments determine favorable
configurations within an equal-energy manifold.

When the density of electrons $\rho$ is larger, a sub-extensive number of states, possibly even a unique one, tend to be favored, because they manage to
minimize the kinetic energy; this we term the energetic regime.

In Section \ref{sec_lowdoping}, we present entropic considerations relevant to
the regime of low doping.
The total entropy contribution comes from both the loops (with parameters extracted
numerically), and the electrons (derived analytically).
In particular, the presence of electrons acts like a cutoff on the total number of loops in the system. In 3D, forbidding configurations with less loops than electrons suppresses the formation
of extensive loops (present in the peak of the probability distribution function (PDF), see e.g. figure \ref{fig:pdfelec3d}), but does not modify the exponents of the PDF. On the other hand in 2D, there is evidence for a variation in the exponent of the power law of the loop distribution.

Once the electronic density gets too large for loops to be restricted to at
most one electron, we need to consider energetics.
Section \ref{sec_phasediagram} presents these energy considerations and the
phase diagram obtained thereby.
Ignoring lattice constraints on loop coverings, we use energy calculations and
a Maxwell construction to obtain the phase diagram of Figure \ref{fig:PD}.  We verify some of these results through Monte Carlo simulations.  We also 
present constraints imposed by the lattices under considerations and identify loop crystals arising as the doping is varied.

With periodic boundary conditions, loops spanning the system can be divided into segments connecting ``opposite'' faces, which we call filaments. Because the transmission of electrons through the system can only occur via these conducting channels, section \ref{sec:conduction} is dedicated to their statistics, as a function of doping $\rho$ and dimension.
In absence of itinerant electrons, the number of filaments grows linearly with (cubic) system size in 3D but remains constant and of $\mathcal{O}(1)$ in 2D. While these behaviors are qualitatively not modified at low doping, the conducting channels turn out to vanish at intermediate values of $\rho$.

A separate final section is devoted to an outlook which also contains some
words on the behaviour of analogous models on the triangular and kagome
lattices, which turn out to exhibit qualitatively different properties both
from the pyrochlores and from each other: we find a magnetic conducting solid as well as an insulating cooperative paramagnet.

The numerical component of our work involves Monte Carlo simulations of
several types; some details are provided in the Appendix.

\section{The system and the loop framework} \label{sec:loop_framework}

\subsection{The model}

\begin{figure}[tbp]
\centering\includegraphics[width=0.8\columnwidth]{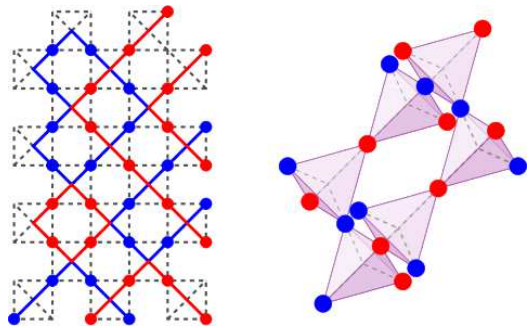}
\caption{
In the Coulomb phase, each frustrated unit (crossed squares in 2D
checkerboard, left, and tetrahedra in 3D pyrochlore, right) possesses two up
and two down spins, respectively colored in blue and red. Those are the so-called ice-rules or divergence free conditions. Connecting spins of
the same color forms a network of loops, as illustrated for the checkerboard.
This model is equivalent to the nearest neighbour spin ice
model.}
\label{fig:latt}
\end{figure}

We will focus primarily on the checkerboard and pyrochlore lattices, which are
two lattices where a Coulomb phase can appear (see figure~\ref{fig:latt}). The
localized magnetic moments are Ising spins $\mathbf{S}_{i}$ all parallel to a
\textit{global} axis, while itinerant electrons can hop on the lattice
sites. The Hamiltonian is
\begin{eqnarray}
\mathcal{H} &=&~ J\sum_{\left<i,j\right>} \mathbf{S}_{i}\cdot\mathbf{S}_{j} 
~-~ \sum_{\left<i,j\right>,\alpha}\;
t\,(c_{i,\alpha}^{\dagger}c_{j,\alpha}\;+\;c_{j,\alpha}^{\dagger}c_{i,\alpha})  \nonumber
\\
&-&~ J_{h}\sum_{i,\alpha,\beta} c_{i,\alpha}^{\dagger}(\mathbf{\sigma}_{\alpha,\beta}\cdot\mathbf{S}_{i})\;c_{i,\beta} 
\label{eq:ham1}
\end{eqnarray}
where $t$ is the hopping integral between two neighboring sites,
$c_{i,\alpha}^{\dagger}$ ($c_{i,\alpha}$) are creation (annihilation)
operators of itinerant electrons of spin $\alpha$ on site $i$, and
$\sigma_{\alpha,\beta}$ are the Pauli matrices.
In order of appearance, the terms in equation~(\ref{eq:ham1}) are the
antiferromagnetic nearest-neighbour exchange between the localized spins
incorporating magnetic frustration, the hopping term allowing movement of itinerant
electrons, whose spins interact with the localized magnetic moments
through ferromagnetic Hund coupling (last term). 

In this work, we focus on the limit $t\ll J_{h}\ll J$. In this limit, the
highly degenerate ground state of the frustrated system serves as background
for the motion of the electrons.  Magnetic excitations (violations of ice
rules) are not present in this limit.  Electrons can only hop between nearest
neighbour spins having the same orientation.

At zero temperature, a N\'eel or ferromagnetic order would give rise to an
insulating or metallic state respectively, but a spin liquid provides a
network of conducting paths for itinerant electrons.
The present work unveils the geometry of this network, as it is influenced by
both the lattice and the minimization of the hopping energy.

\subsection{Emergence of loops}

The pyrochlore and checkerboard lattices are made of corner-sharing units
with four spins; respectively the tetrahedron and the square with crossings.
The antiferromagnetic couplings impose the so-called ``ice-rules'' with zero
magnetization per unit, obtained with two spins pointing up and two spins
pointing down~\cite{Bernal33a}.  In absence of itinerant electrons, this
ground state is highly degenerate with 6 possible configurations per unit. It
corresponds to the 6-vertex model in 2 dimensions~\cite{Baxter07a} and can be
mapped onto the nearest neighbor spin ice model in 3
dimensions~\cite{Harris97a,Moessner98b}.  These spin systems serve as background for the
emergent physics of the so-called \textit{Coulomb phase}~\cite{Henley10a}, a
gauge theory where the discrete ice-rules under coarse-grained lead to the emergence of a divergence free flux. 

Joining spins of the same orientation in every unit, one obtains loops of
up spins and loops of down spins (see Fig.~\ref{fig:latt}).  The Coulomb phase can thus be
described as an ensemble of possible loop coverings.  The resulting loop model
possesses two flavors (loops of up and down spins), where every site of the
premedial lattice~\cite{Henley10a} is occupied by two loops, one of each
flavor, and every bond is visited by one loop only.
In previous work by some of the authors~\cite{Jaubert11a}, a detailed account
has been given of the statistics and distribution of these loops, both for the
2D and 3D cases.  

In the limit of large $J_h$, all up (down) electrons are only allowed to hop
along an up (down) loop, and are constrained to remain within this loop.  This
reduces the electron dynamics to be one-dimensional whatever the
dimension of the lattice.

The 1D hopping restriction allows us to describe electron dynamics in terms of
the dispersion of a 1D tight-binding problem, $E_k=-2t\cos{k}$, with $k$ the
1D momentum along the loop.  The 1D momentum along the loop, $k=2\pi q/\ell$,
is discrete for a loop of finite length $\ell$ ($q = -\ell/2, ... , \ell/2-1$).  An up (down) loop of length
$\ell$ can contain between 0 to $\ell$ up (down) electrons.  Note that double
occupancy does not occur in the limit we are considering.

\begin{figure}[tbp]
\centering\includegraphics[width=0.99\columnwidth]{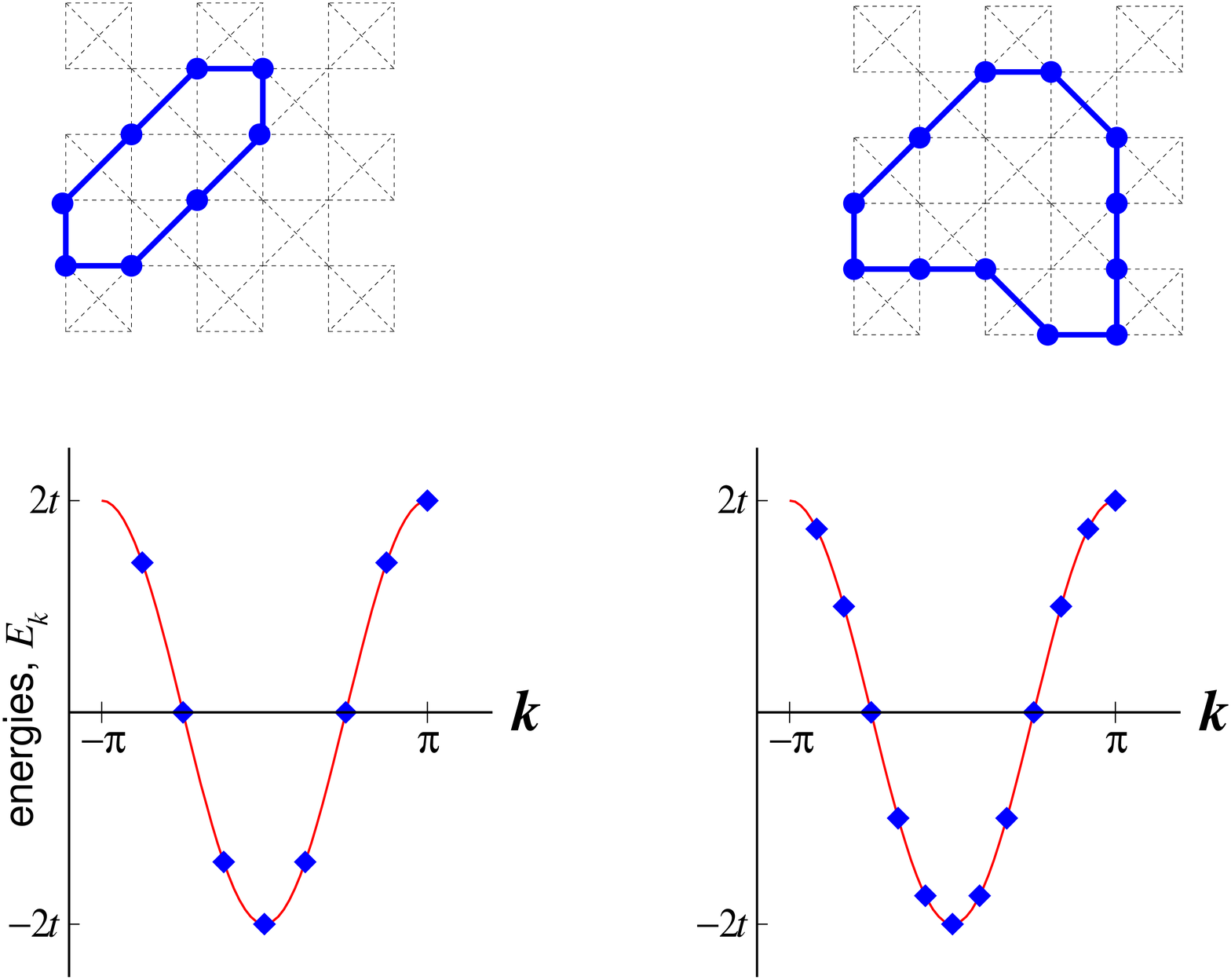}
\caption{
Single-particle energy levels of itinerant electrons confined to loops of
length $\ell=8$ (left) and $\ell=12$ (right), due to the hopping term in
Hamiltonian~\eqref{eq:ham1}. There are nondegenerate levels at the highest and
lowest energies, $\pm{2t}$, independent of loop length; the other levels are
doubly degenerate.
}
\label{fig:nrjlevel}
\end{figure}

The lowest and highest single-particle levels in the dispersion have energy
$\pm 2t$, independent of the loop length (figure~\ref{fig:nrjlevel}).

Our system is thus described by 

\begin{itemize}
\item  the number of lattice sites $N$, equal to $4\,L^{2}$ and $16\,L^{3}$
for the checkerboard and pyrochlore lattices respectively, where $L$ is the
linear number of unit cells;
\item the total electron number $N_e$, and the electron density  $\rho=N_e/N$;
\item the loop histogram of a given configuration, \textit{i.e.}  the number $h_{i}$ of
loops of length $\ell_{i}$, $\ell_{i}$ being necessarily an even number on a
bipartite lattice;
\item the type of lattice, which will among other things determine the
smallest possible loop in the system $\ell_{min}$ (4 for checkerboard and 6
for pyrochlore); the longest possible loop length is always $\sim N/2$.
\end{itemize}

We define a few additional relevant observables:

\begin{itemize}
\item the total number of loops in a given configuration $N_{\ell}=\sum_{i} h_{i}$;
\item the average loop length for a given configuration\vspace{0.2cm}\\
$\overline{\ell}\;=\;\dfrac{\sum_{i} h_{i} \ell_{i}}{\sum_{i} h_{i}}\;=\;\dfrac{N}{N_{\ell}}$;
\item the statistical average loop length $\langle \ell \rangle$ over all
loop configurations;
\item the statistical average number of loops $\langle N_{\ell} \rangle$;
\item the number of filaments (see section~\ref{sec:conduction}).
\end{itemize}


\section{Low doping regime} \label{sec_lowdoping}

In this section we consider the low doping regime where $N_e$ is small enough
to have loop configurations with more loops than electrons, $N_{\ell}>N_e$.
Since the lowest single-electron energy level in any loop is $-2t$, the
minimum accessible energy is the same ($-2tN_e$) for all such configurations.
Therefore, the ground state manifold consists of all such loop coverings
with the same energy $-2tN_e$.  The free energy within this manifold is then only determined by
entropics.  At zero temperature, entropy is understood in the
sense that all configurations have an equal probability to occur.

In the first subsection below, we present some entropy calculations, combining
loop and electronic contributions to the entropy, and show how this determines
the average loop length at nonzero electronic density $\rho$.  In the second
subsection we present numerical results on the effect of electrons on the
entire loop length distribution (loop PDF).  The effect on the loop PDF is a
natural way to characterize the influence of electrons in the magnetic
Coulomb-phase system.

\begin{figure*}
\centering\includegraphics[width=0.9\columnwidth]{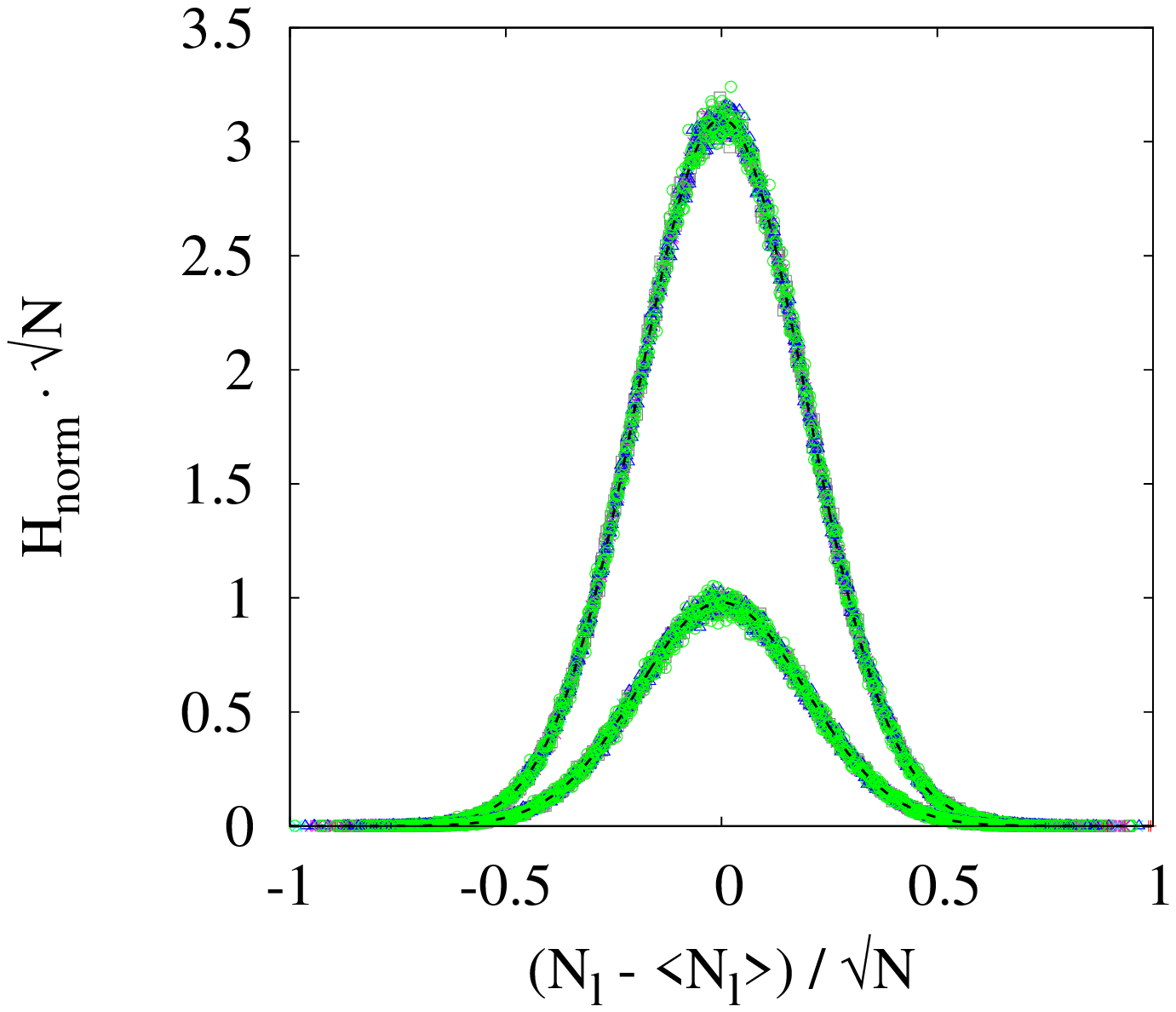}
\centering\includegraphics[width=0.9\columnwidth]{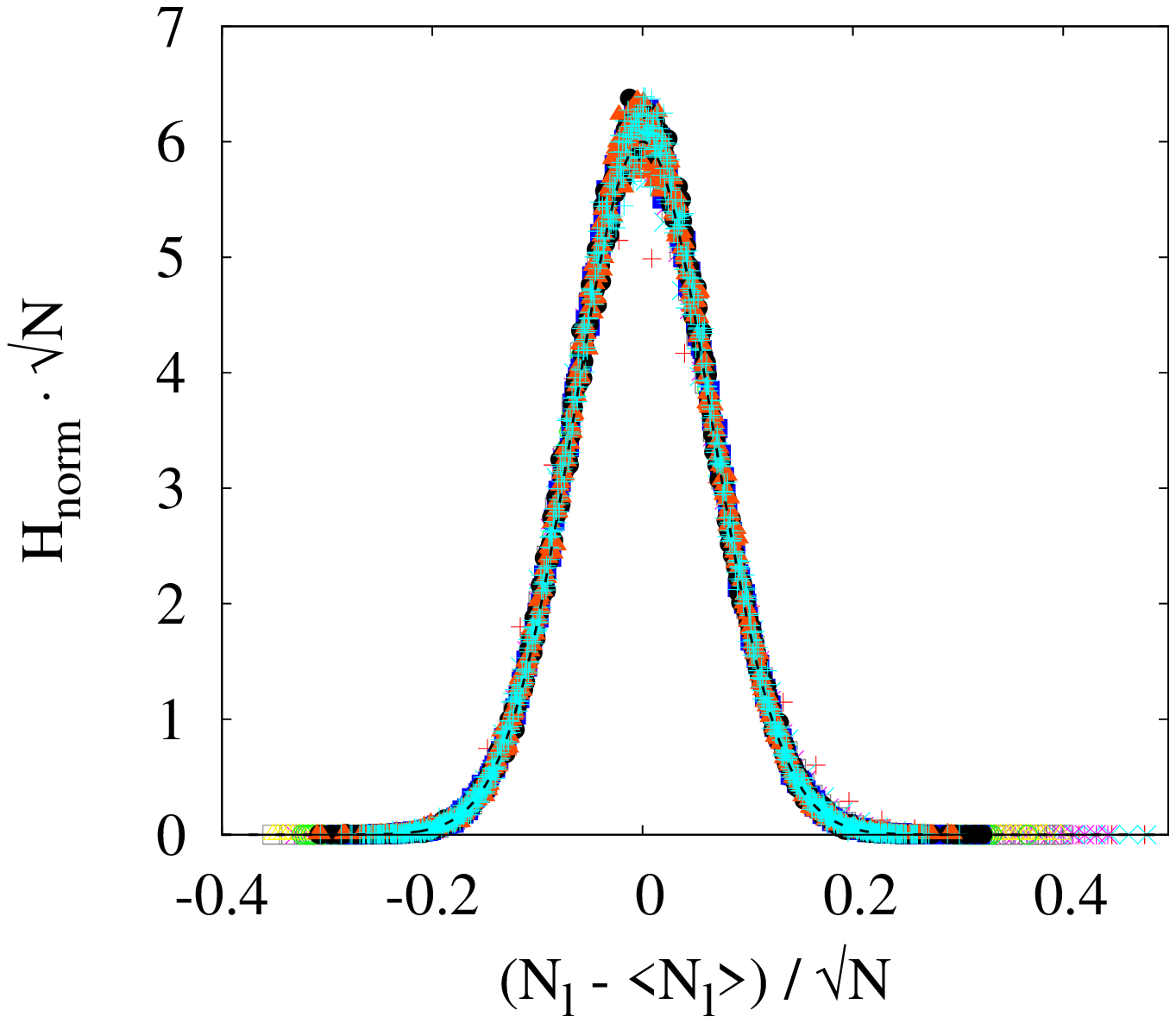}
\caption{%
Histogram of distribution of the number of loops $N_{\ell}$ per configuration
in 2D (left) and 3D (right), in the absence of electrons. The distribution is
scaled according to the Gaussian expression of equation~(\ref{eq:Gaussian}) 
collapsing all system
sizes onto the same curve ($L=100$ to 600 in 2D and $L=4$ to 60 in 3D).  In
2D, the two Gaussians have the same width and corresponds to even and odd
$N_{\ell}$ (upper and lower curve).  In 3D, finite size corrections are visible
for $L=4$ (red crosses).
}
\label{fig:entropy2}
\end{figure*}

\subsection{Entropy}

The total entropy consists of loop and electronic contributions.  The loop
distribution is of course itself affected by the itinerant electrons.
However, in the limit of small $\rho$, the change in loop entropy is small.  Below we combine the $\rho=0$ loop entropy with the finite-$\rho$
electronic entropy to approximate the total entropy at small $\rho$.

\paragraph*{Loop entropy.}

Figure 3 shows the distribution of the number of loops $N_{\ell}$ having
length $\ell$, in the absence of electrons ($\rho=0$), obtained from Monte
Carlo simulations (Appendix \ref{appworm}).  The distribution has Gaussian
form:
\begin{eqnarray}
P(N_{\ell})\sim \dfrac{1}{\sqrt{N}}\;\exp\left[-\dfrac{(N_{\ell}-\langle N_{\ell}\rangle)^{2}}{2\kappa N}\right]
\label{eq:Gaussian}
\end{eqnarray}
where $\kappa_{2d}=0.0384$ and $\kappa_{3d}=0.00423$. In 2D, there
are \textit{two} Gaussians, corresponding to even and odd $N_{\ell}$.

For a system with short-distance correlations, Gaussian distributions are
natural to expect from the Central Limit Theorem, since by dividing the system
into small mesoscopic segments the total distribution can be recast into a sum
of many random variables.  In our case, however, we have a system with
algebraic correlations and extended objects (loops), so finding Gaussian
distributions is not a priori trivial.

Equation~\ref{eq:Gaussian} can be expressed in terms of the average loop length $\overline{\ell}=N/N_{\ell}$
instead of $N_{\ell}$:
\begin{multline}
P(\overline{\ell})\equiv P(N_{\ell}) \left|\dfrac{{\rm d}N_{\ell}}{{\rm
d}\overline{\ell}}\right|  \, 
\sim \dfrac{\sqrt{N}}{\overline{\ell}^{2}}\;\exp\left[-\dfrac{N}{2\kappa}\left(\frac{1}{\overline{\ell}}-\frac{1}{\langle \ell\rangle}\right)^{2}\right]
\\
\sim \dfrac{\sqrt{N}}{\overline{\ell}^{2}}\;\exp\left[-\dfrac{N}{2\kappa}
\left(\frac{\overline{\ell}-\langle \ell\rangle}{\langle \ell\rangle^{2}}\right)^{2}\right]
\label{eq:Gaussian2}
\end{multline}
Here we have used $1/\overline{\ell}\langle\ell\rangle\approx
1/\langle\ell\rangle^2$, which is valid for large $N$ in the region where the
Gaussian is appreciable.  Thus the loop contribution to the entropy
($\sim\ln{P}$) is
\begin{equation}
S_{loop}=S_{1}-\dfrac{N}{2\kappa\,\langle \ell\rangle^{4}}\left(\overline{\ell}-\langle \ell\rangle\right)^{2}
\, ,
\label{eq:entropy2}
\end{equation}
where $S_1$ is a constant.

\paragraph*{Electronic entropy.}

Since we have at most one electron per loop, the number of possible
combinations to put $N_{e}$ electrons in $N_{\ell}$ loops is the binomial
(Pascal) coefficient $N_l!/[N_e!(N_l-N_e)!]$.  The logarithm then gives the
electronic contribution to the entropy.  Using Stirling's approximation for
the thermodynamic limit ($N_l\gg1$, $N_e\gg1$), we get as per usual
\begin{equation}
S_{elec} =  \frac{N}{\overline{\ell}} \left[x \ln x + (1-x) \ln(1-x)\right]
\end{equation}
where $x= N_e/N_l = \rho \overline{\ell}$.  
%

\begin{figure*}
\centering\includegraphics[width=0.9\columnwidth]{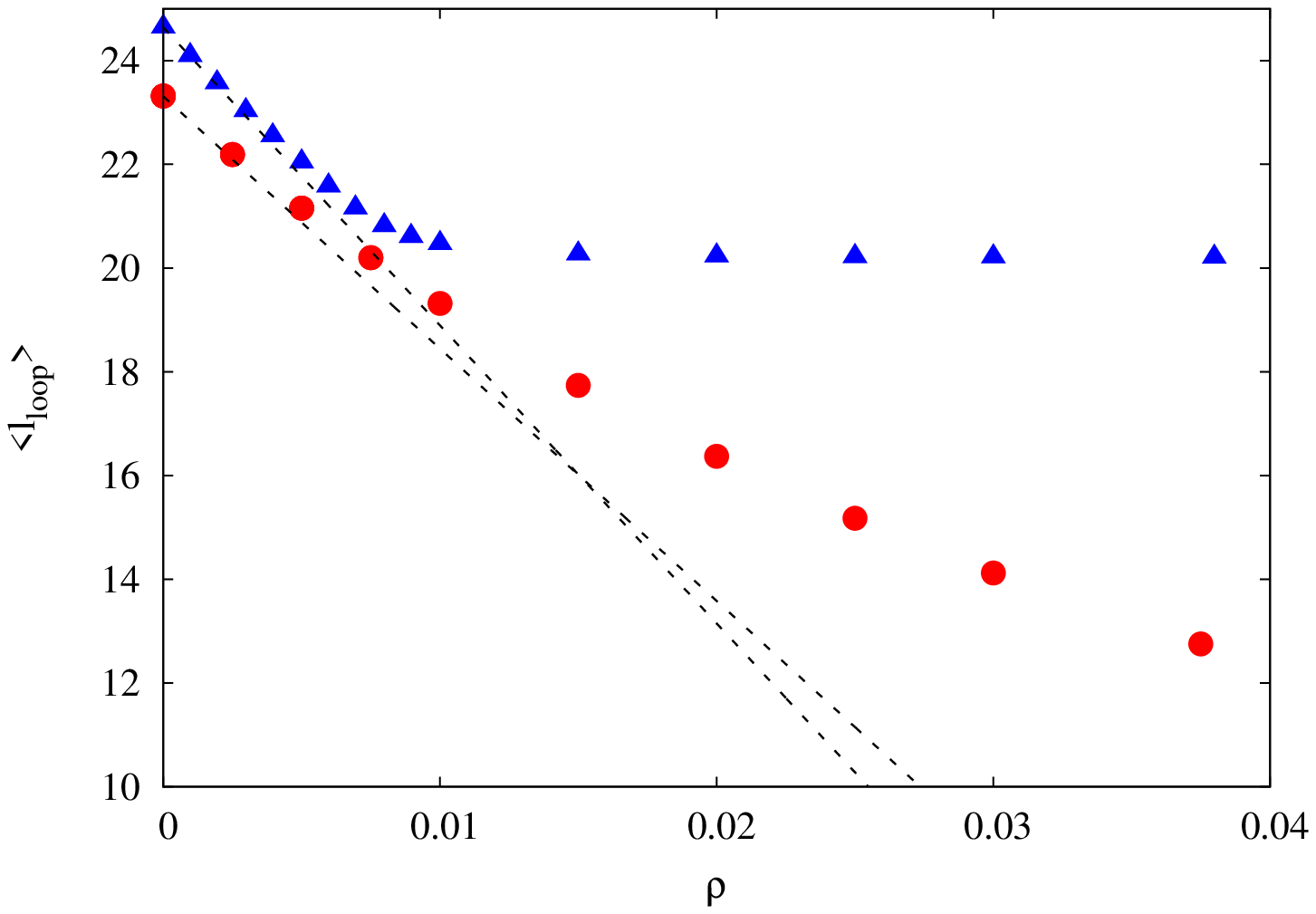}
\centering\includegraphics[width=0.9\columnwidth]{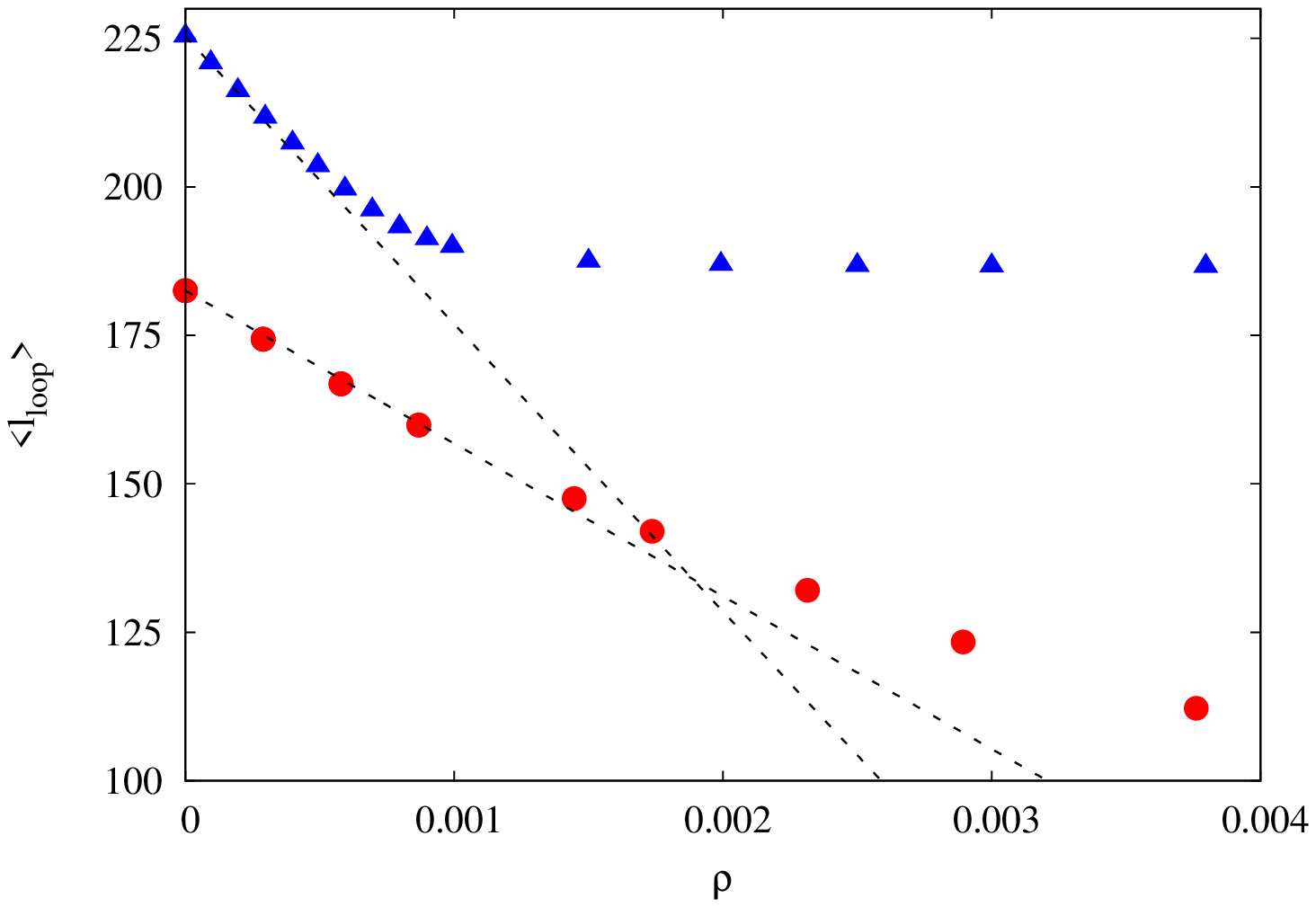}
\caption{
Mean value of the loop length $\langle\ell\rangle$ as a function of
doping.  
\textit{Left}: checkerboard, $L=10$ (red dots) and 60
(blue triangles). 
\textit{Right}: pyrochlore, $L=6$ (red dots) and 20
(blue triangles). 
The dashed straight lines are predictions from Equation \eqref{eq:ell0},
without any fitting parameters. For both panels, the blue triangles data points are very close to thermodynamic limit.
}
\label{fig:meanell_rho}
\end{figure*}

\paragraph*{Total Entropy.}

For small electron densities, we can approximate the loop entropy by the
$\rho=0$ expression calculated above.  The total entropy of electrons and
loops is then
\begin{multline}
S_{tot}=\;-\; \frac{N}{\overline{\ell}} \left[\rho \overline{\ell} \ln \rho \overline{\ell}
+ (1-\rho \overline{\ell}) \ln(1-\rho \overline{\ell})\right] \\
~+~
S_{1}-\dfrac{N}{2\kappa\,\langle \ell\rangle_0^{4}}\left(\overline{\ell}-\langle \ell\rangle_0\right)^{2}
\, .
\label{eq:Selecloop}
\end{multline}
The subscript 0 represents the ensemble average at $\rho=0$.  
The value of $\overline{\ell}$ that maximizes $S_{tot}$ is found (in the limit
$\rho\overline{\ell}\ll1$) to be
\begin{equation}
\overline{\ell}_{opt} \approx \langle \ell\rangle_0 \;-\; \rho\kappa\langle \ell\rangle_0^{3} \, .
\label{eq:ell0}
\end{equation}
For $\rho \langle \ell\rangle_{0} \ll1$, the distribution stays almost Gaussian, so this most probable value of $\overline{\ell}$ is approximately the mean value of the distribution
$\langle\ell\rangle_{\rho}$.  In Figure \ref{fig:meanell_rho}, we compare this
prediction to numerical data.  For each system size $L$, we can extract
$\langle\ell\rangle_0$ from the numerical $\langle\ell\rangle$ at $\rho=0$.
Eq.~\eqref{eq:ell0} then gives a linear prediction (dashed straight lines in
Figure \ref{fig:meanell_rho}), which works well for small $\rho$.

\subsection{Loop length distribution}

We next examine the effect of electrons on the entire loop probability
distribution function (PDF).  We denote by $\tau$ the power-law exponent, when
the PDF has form $\ell^{-\tau}$.  Without electrons ($\rho=0$), the PDF
follows $P_{2D}{\sim}L^{2}/\ell^{\tau}$ in 2D with $\tau=2+1/7$.  In 3D, the
$\rho=0$ PDF displays a crossover around $\ell\sim L^{2}$ between two power
laws, from $L^{3}/\ell^{5/2}$ to $1/\ell$.  (For details, see
Ref.~\onlinecite{Jaubert11a}.)

\begin{figure}
\centering\includegraphics[width=\columnwidth]{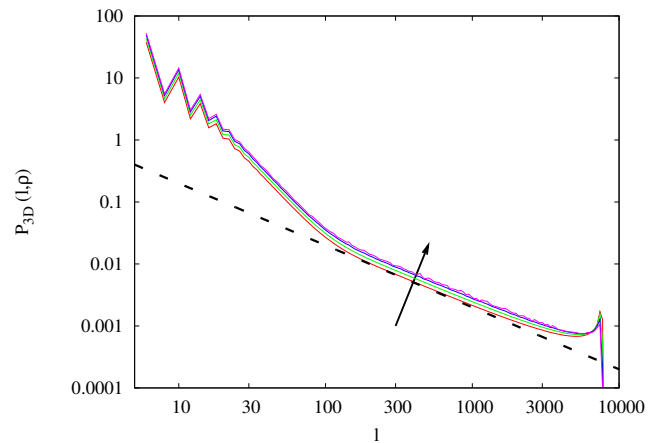}
\caption{
Loop length distribution $P_{3D}(\ell,\rho)$ in 3D for different values of doping ($\rho = \{0, 0.00075, 0.0015, 0.002\}]$) and system size $L=10$. The distribution is normalized such that $\int \ell P_{3D}(\ell,\rho) \textrm{d}\ell =16 L^{3}= N$.
 The dashed line indicates the power law fit $\ell^{-1}$ at $\rho=0$ and the arrow shows the shifting of the distribution for increasing $\rho$. The exponents of the two power law regions (before and after $\ell\approx L^{2}= 100$) do not vary, but the peak for long winding loops $\ell \sim 8 L^{3}=N/2$ gets smaller with increasing $\rho$.
 }
\label{fig:pdfelec3d}
\end{figure}

\begin{figure}[ht]
\centering\includegraphics[width=\columnwidth]{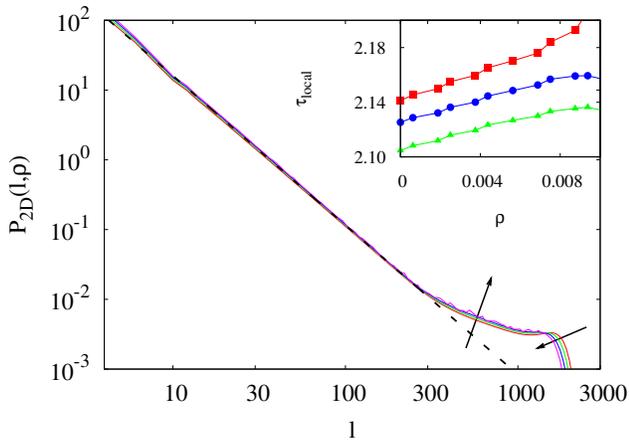}
\caption{
\textit{Main:} Loop length distribution $P_{2D}(\ell,\rho)$ in 2D for different values of doping ($\rho = \{0, 0.0025, 0.0056, 0.0087\}]$) and system size $L=40$. The distribution is normalized such that $\int \ell P_{2D}(\ell,\rho) \textrm{d}\ell =4 L^{2}= N$.
 The dashed line indicates power law $L^{2}/\ell^{15/7}$ and the arrow shows the shifting of the distribution for increasing $\rho$. On this scale, it is not obvious whether the exponent of the power law varies or not.
\textit{Inset:} The power-law exponent defined locally on $\ell$,  $\tau_{\rm local}$
for $\ell=8 (\textcolor{red}{\blacksquare}), 16 (\textcolor{green}{\blacktriangle}), 32 (\textcolor{blue}{\bullet})$ for $L=40$ (similar behavior obtained for $L=20$ and 60).
}
\label{fig:pdfelec2d}
\end{figure}

At finite electron densities, loop configurations with $N_{\ell}<N_{e}$ are
rejected due to energetics, as explained previously.  Thus electron doping
acts not entirely unlike a ``chemical potential'' for loops, favoring
configurations with more loops, and thus a priori shorter ones.  In both 2D and 3D, this implies a
disappearance of longer loops, as can be seen in the calculated PDF's of
Figures \ref{fig:pdfelec3d} and \ref{fig:pdfelec2d}, where the large-$\ell$
parts of the PDF are progressively decimated for increasing $\rho$.  In 3D the
form of the PDF is otherwise unchanged (Figure \ref{fig:pdfelec3d}).  In 2D
the effect seems to be more drastic; the inset of figure \ref{fig:pdfelec2d} suggests that the entire power-law behavior of the PDF is modified.

To quantify how the 2D loop PDF changes qualitatively at finite $\rho$, we
define and compute a \textit{local} exponent in $\ell$: $\tau_{\rm
local}(\ell,\rho)=\log\left(P_{2D}(\ell,\rho)/P_{2D}(2\ell,\rho)\right)/\log
2$.  This is displayed in the upper right inset to Figure \ref{fig:pdfelec2d}. Our results suggest a trend toward increasing $\tau_{\rm local}$ as a function of $\rho$, consistent
with the idea that itinerant electrons favor small loops.

This outcome deserves a few comments. In Ref.~\onlinecite{Jaubert11a}, the
loop statistics of the 2D Coulomb phase (zero doping) has been shown to be
analogous to the Stochastic Loewner evolution process SLE$_{\kappa=6}$ with
fractal dimension $D_{f}=1+\kappa/8=7/4$.  The SLE$_{\kappa}$ can be
identified to various realizations of the $\mathcal{O}(n)$ model through the
relation $n = -2 \cos(4\pi/\kappa)$.~\cite{Cardy05a} The $\mathcal{O}(n)$
model is often used to describe fully packed loop models with loop fugacity
$n$ (see \textit{e.g.}  Ref.~\onlinecite{Baxter07a}). The partition function of the fully
packed loop model is $\mathcal{Z}=\sum n^{N_{\ell}}$, where the sum runs over
all possible configurations. The Coulomb phase corresponds to a fully packed
loop model~\cite{Jaubert11a}; at zero doping, the free energy of our model is
trivially independent of the number of loops $N_{\ell}$ and thus corresponds
to a fugacity $n=1$. Higher values of the fugacities favor configurations with
more loops and tend to increase the value of $\tau$~\cite{Jacobsen99a}, in a
way reminiscent of the influence of doping here. The addition of itinerant
electrons remains a non-trivial problem and is not exactly the same as a
fugacity for a loop, but at small and finite doping $\rho$, some features
could be captured by $\mathcal{O}(n(\rho)>1)$ models or SLE$_{\kappa(\rho)<6}$
processes.


\section{Large densities; phase diagrams} \label{sec_phasediagram}

In this section, we present results relevant for higher densities, where
entropic considerations are no longer sufficient, and non-trivial electronic hopping energies need to be
considered.  We continue to describe the system in terms of loop coverings.

We first provide an analysis based on calculations for loop coverings of equal-length loops.  Using a Maxwell construction, we can use this information to predict ranges of electron
density where the ground state manifold consists of coverings by loops of two
different lengths.

These considerations, described in the first two subsections below, do not
take into account any lattice constraints other than the fact that the minimum
loop length is $\ell_{min}=4$(6) for the checkerboard (pyrochlore).  Lattice
constraints, disallowing some loop coverings, are difficult to enumerate or
list comprehensively on account of their non-local nature.  In the final subsection, we present the lattice constraints that we have identified, and their implications.

\subsection{Equal-length loop configurations}

In this subsection and the next, we imagine that coverings with any unique loop
length $\ell \geqslant \ell_{min}$ are possible.  
Later, we will show that at certain fillings, the ground states exhibit a unique loop length, while at others, the behaviour can be understood through a Maxwell construction based on the
single-length results.

For an odd number of electrons $\rho \ell=2n_{o}+1$ in a loop of length $\ell$, the energy is (see
figure~\ref{fig:nrjlevel})
\begin{eqnarray}
E(\rho,\ell)=-2t\;\sum_{n=-n_{o}}^{n_{o}}\;\cos\left(\dfrac{2\pi n}{\ell}\right)= -2t \; \dfrac{\sin\left(\pi\rho\right)}{\sin\left(\pi/\ell\right)}
\label{eq:nrjloopodd}
\end{eqnarray}
For an even number of electrons $2n_{o}$, this expression becomes
\begin{eqnarray}
E(\rho,\ell)&=& -2t \; \dfrac{\sin\pi\left(\rho-1/\ell\right)}{\sin\left(\pi/\ell\right)} -2t\;\cos(\pi \rho)
\label{eq:nrjloopeven}
\end{eqnarray}
More generally, if $\rho\ell$ is not an integer, we define the highest odd
integer below $\rho\ell$ as 
\begin{eqnarray}
\eta\;=\;2 \textrm{E}\left(\dfrac{\rho\;\ell-1}{2}\right)+1
\label{eq:eta}
\end{eqnarray}
where E$(.)$ is the floor function.  Each loop is filled with at least $\eta$
electrons up to the energy levels at $k=\pm\pi(\eta-1)/\ell$, while the
remaining $N(\rho-\eta/\ell)$ electrons in the system are distributed in the
partially filled level at $k=\pm\pi(\eta+1)/\ell$.  The total energy is
\begin{eqnarray}
E(\rho,\ell)= -2t
N \left[ \dfrac{\sin\left(\pi\eta/\ell\right)}{\ell\sin\left(\pi/\ell\right)}
+\left(\rho-\dfrac{\eta}{\ell}\right) \cos\left(\pi \dfrac{\eta+1}{\ell}\right)\right]
\label{eq:nrjloop}
\end{eqnarray}
These expressions are electron-hole symmetric, \textit{i.e.}  invariant under
$\rho\leftrightarrow(1-\rho)$.  

\begin{figure*}[ht]
\centering\includegraphics*[width=\columnwidth]{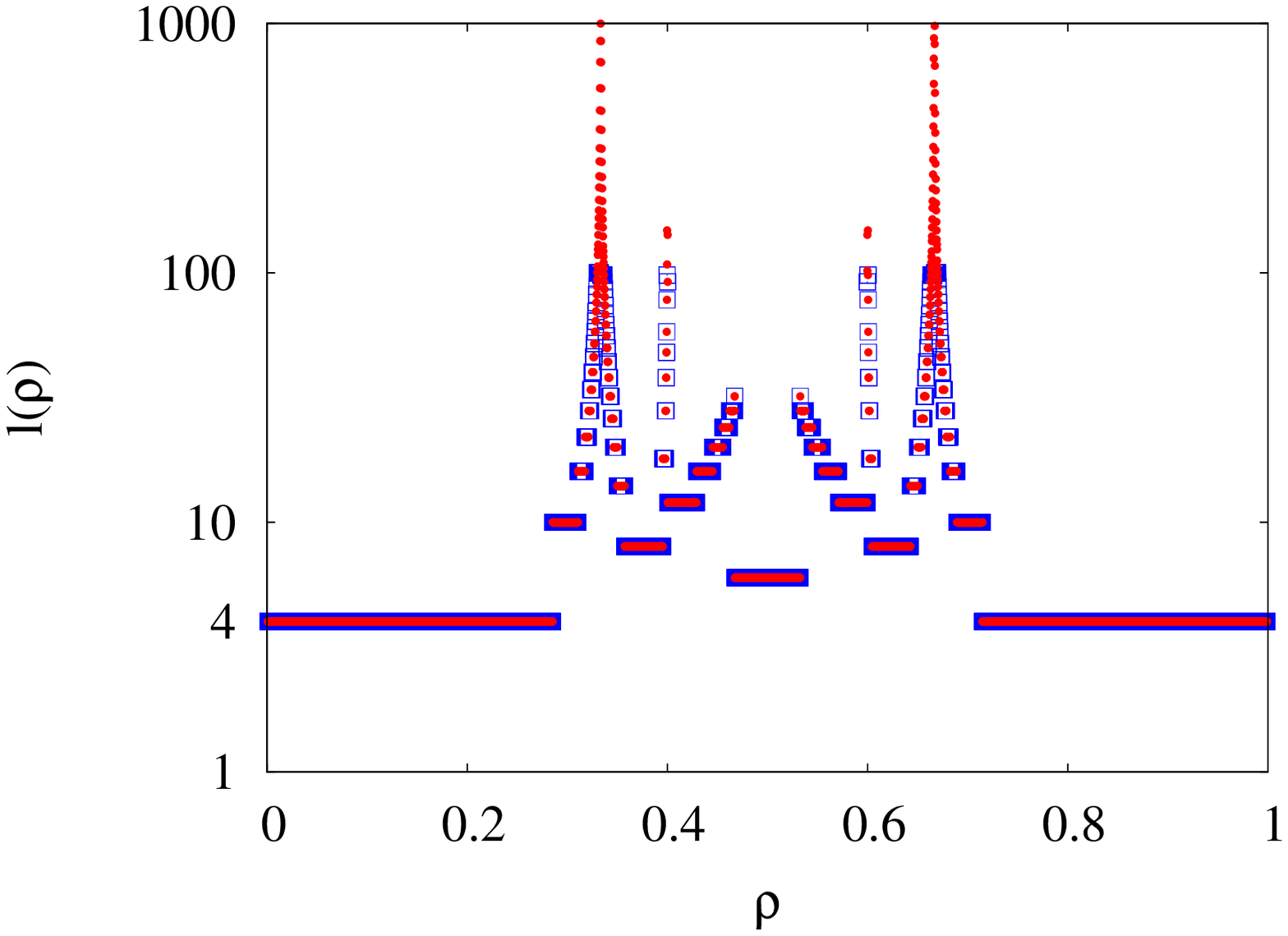}
\includegraphics*[width=\columnwidth]{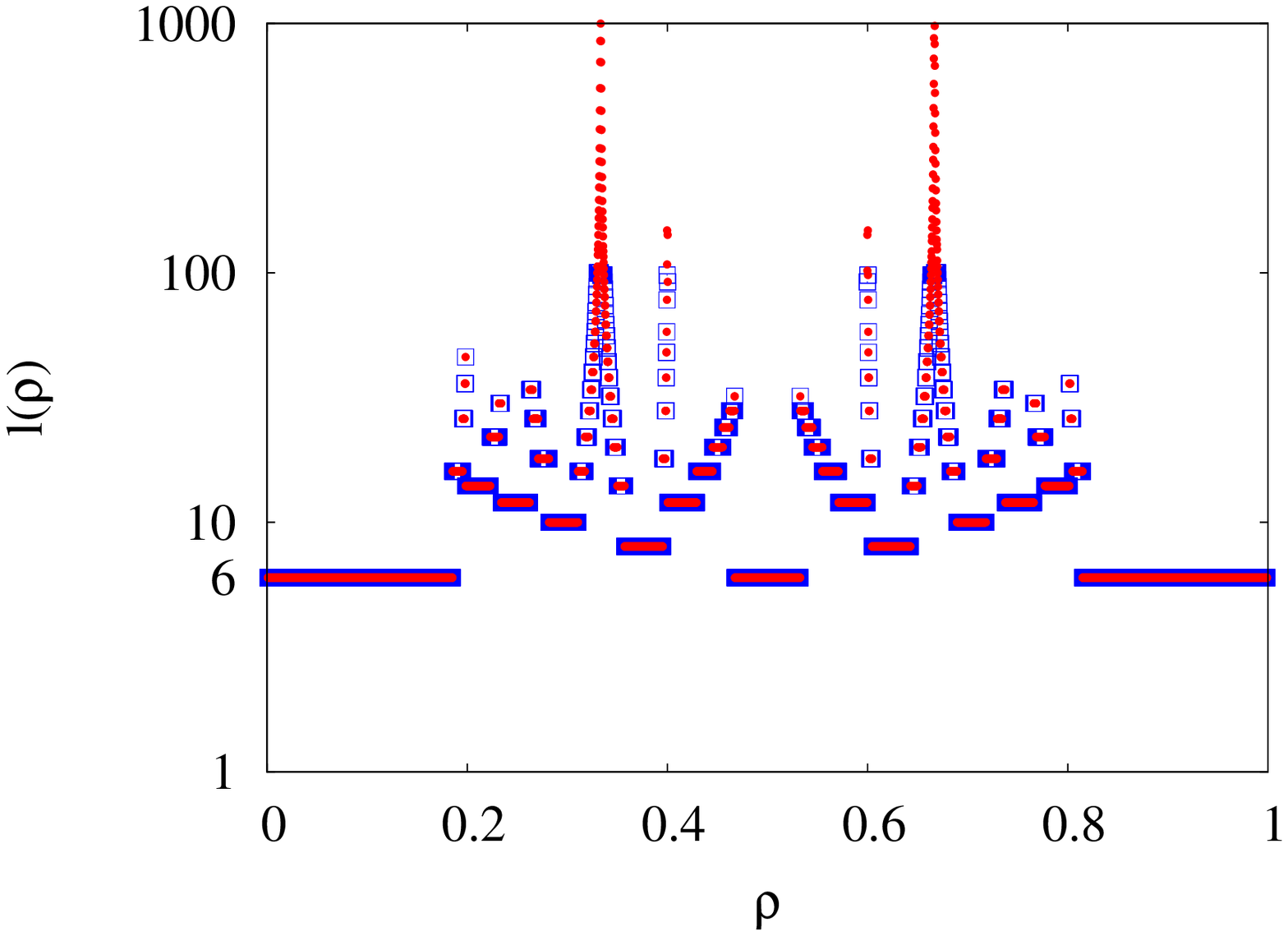}
\caption{%
Loop length $\ell(\rho)$ minimizing the energy~(\ref{eq:nrjloop}) for a system
filled with loops of length $\ell$ only, assuming no lattice constraints other
than $l_{min}=4$ (left panel, checkerboard) or 6 (right panel, pyrochlore).
$\rho$ takes all rational values $p/q$, with $q\in \{0,1,2,...,1000\}$ and
$p\in \{0,1,2,...,q\}$,
and we consider all loop lengths from $\ell_{min}$ to $\ell_{max}=100$ (red
crosses) and 1000 (blue squares): plotting two different values of
$\ell_{max}$ provides a graphical way to visualize the two values of $\rho$
where the most favorable loop length is infinite, namely $\rho=1/3$ and $2/5$
(and their symmetric images with respect to $\rho=1/2$). If several loop lengths give the
same energy, we plot the smallest one.
}
\label{fig:ellrho}
\end{figure*}

Figure~\ref{fig:ellrho} displays the loop length $\ell(\rho)$ that minimizes
the energy~(\ref{eq:nrjloop}) as a function of $\rho$, and thus corresponds to
the ground state if we impose a unique loop length in the system. (The
electron-hole symmetry shows up clearly through the mirror symmetry on either
side of $\rho=0.5$.  Therefore, we shall from now on only consider densities
below 1/2.)

Since the lowest possible energy for an electron is $-2t$ and is only
accessible for one electron per loop, loops of length 4 are favored up to
$\rho=1/4$. Of course for lower densities than 1/4, other configurations may
be possible as long as the number of loops is larger than the number of
electrons, but a system with only loops of length 4 will always be part of the
ground state manifold for $\rho \leqslant 1/4$.

Remarkably, for most densities, a finite loop length with discrete energy
levels is preferred compared to infinite loops, except for $\rho=1/3$ and
$2/5$.

We plot on figure~\ref{fig:nrjrho} the minimum energy corresponding to the
loop length $\ell(\rho)$ of figure~\ref{fig:ellrho}:
$E_{min}(\rho)=E(\rho,\ell(\rho))$.

\subsection{Maxwell construction; lattice-independent phase diagram}

We now move beyond configurations with unique loop length.
We need to consider mixtures of electron densities in the different
loops.
This is done through a Maxwell construction, similar to the physics of a
liquid-gas first-order phase transition.

\begin{figure*}[ht]
\centering\includegraphics[width=\columnwidth]{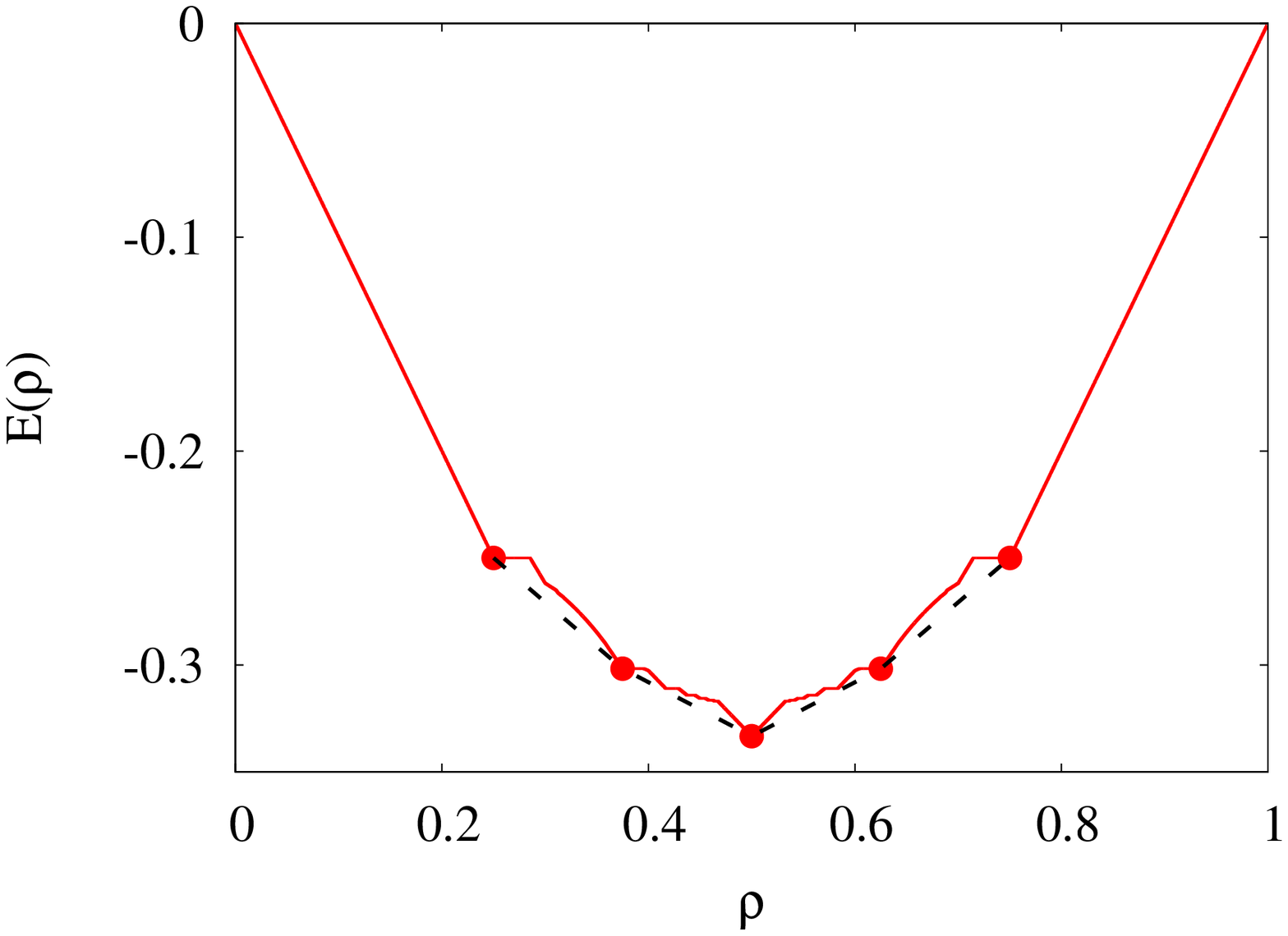}
\includegraphics[width=\columnwidth]{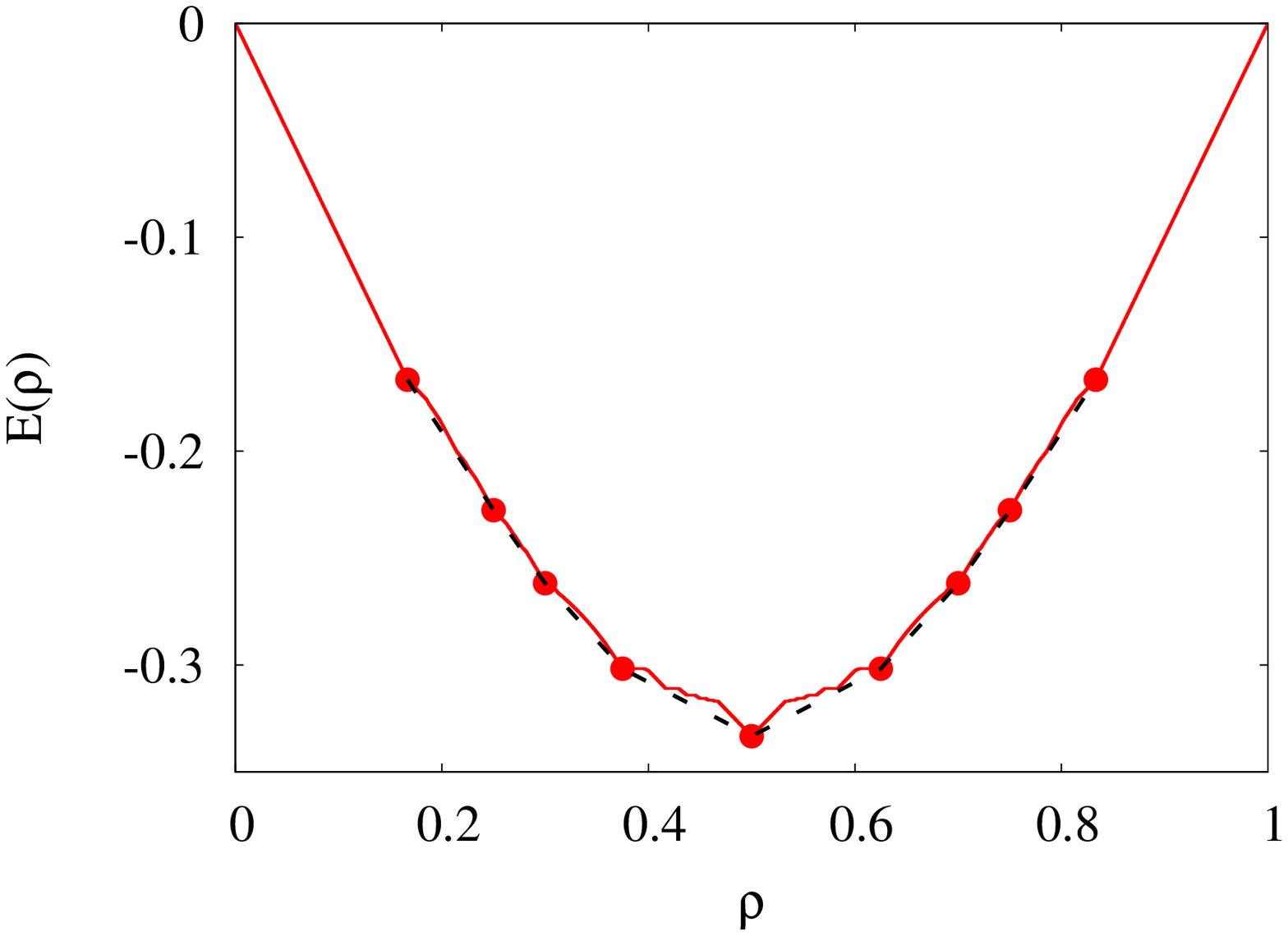}
\caption{%
Minimum energy $E(\rho)$ corresponding to the value of $\ell(\rho)$ plotted on
figure~\ref{fig:ellrho}, for a system filled with loops of length $\ell$ only,
assuming no lattice constraints other than $l_{min}=4$ (left, checkerboard) or
6 (right, pyrochlore). $\rho$ takes all rational values $p/q$, with
$q\in \{0,1,2,...,1000\}$ and $p\in \{0,1,2,...,q\}$.  We consider all loop
length from $\ell_{min}$ to $\ell_{max}=1000$. We chose arbitrarily $t=1/2$.
A Maxwell construction can be visualized from this figure. The red dots are a
set of points, such that the dashed lines connected them are always below the
red curve $E(\rho)$. Hence a mixture of two phases corresponding to two
consecutive red dots $(\ell_1,\rho_1)$ and $(\ell_2,\rho_2)$ has a lower
energy than a single phase with a unique density of electrons $\rho$ on a
unique type of loop length $\ell$. For increasing values of $\rho$, the red dots correspond to the most favored loop length $\ell=\{4,8,6,8,4\}$ (left panel) and $\ell=\{6,12,10,8,6,8,10,12,6\}$ (right panel), as can be read from figure~\ref{fig:ellrho}.}
\label{fig:nrjrho}
\end{figure*}

A system of $N$ sites and density $\rho$ can be divided into two subsets of
sites $N_{1}$ and $N_{2}$, with density $\rho_{1}$ and $\rho_{2}$ of electrons and loop length $\ell_{1}$ and $\ell_{2}$ respectively, with
\begin{gather}
N\;=\;N_{1}\;+\;N_{2}  \nonumber \\
N\;\rho\;=\;N_{1}\;\rho_{1}\;+\;N_{2}\;\rho_{2}.
\label{eq:csv}
\end{gather}
Now if a straight line between $E_{min}(\rho_{1})$ and $E_{min}(\rho_{2})$
remains below the curve $E_{min}(\rho)$ on figure~\ref{fig:nrjrho}, \textit{i.e.}  if 
\begin{eqnarray}
N\;E_{min}(\rho)\;>\;N_{1}\;E_{min}(\rho_{1})\;+\;N_{2}\;E_{min}(\rho_{2}) \, ,
\end{eqnarray}
then the mixture of two densities is more stable than a unique density and
``phase separation'' occurs. We thus construct a phase
diagram, separating regions of different loop length combinations.

\begin{figure*}[tb]
\centering
\includegraphics[width=1.7\columnwidth]{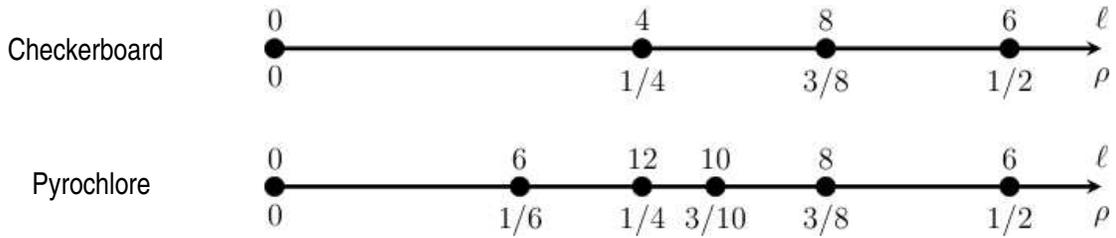}
\caption{
Zero temperature ``phase diagram'' of loop length configuration as a function
of electron density $\rho$, obtained without accounting for lattice
constraints other than $\ell_{min}$=4(6) for checkerboard (pyrochlore).  A
phase is defined by the length of its constituting loops and the density of
electrons on them. The values of $\rho$ and $\ell(\rho)$ for each phase are
given below and above the lines.
}
\label{fig:PD}
\end{figure*}

On figure~\ref{fig:PD}, the dots correspond to densities where a unique loop
length is favored, whereas the zones between them are ``phase mixed'', defined
by the loop lengths and electron densities of the surrounding dots.  The ratio
of one phase compared to the other is given by equations \eqref{eq:csv}. For
example for $\rho=0.4$, 80\% of the sites belong to loops of length $\ell_{1}=8$ with
electron density $\rho_{1}=3/8$, while the remaining 20\% belong to loops of
length $\ell_{2}=6$ with density $\rho_{2}=1/2$. As shown on figure~\ref{fig:ellrho},
loops of length 4 are particularly robust over a wide range of
$\rho$; preventing their formation, \textit{e.g.}  in the pyrochlore lattice in $d=3$ which permits loops of minimal length 6, thus strongly modifies the phase diagram.\\

These results indicate that itinerant electrons tend to favor relatively small loops
and to prevent the formation of infinite ones at zero temperature.

At low electron densities (up to $1/\ell_{min}$), the minimum energy for a
given $\rho$ is degenerate: \textit{e.g.}  for $\rho=1/12$, having only loops of length 4,
6, 8, 10 or 12 gives the same energy, as the number of electrons is smaller
than the number of loops on the system, and every electron can fill the lowest
energy level.  This explains why $E_{min}(\rho)$ is a straight line in this
region (Figure \ref{fig:nrjrho}): as $\rho$ decreases from $1/\ell_{min}$ to
0, the degeneracy of the ground state increases until one recovers naturally
the full degeneracy for $\rho=0$. This of course corresponds to the entropic regime mentioned above.

We stress again that these results have been obtained independently of the
lattice (except for the value of $\ell_{min}$), and that lattice constraints
(next subsection) will modify some of the phase diagram.

\subsection{Lattice constraints on loop coverings}  \label{sec:lattice_constraints}

We now consider effects of the lattice in disallowing some of the
configurations predicted by our analysis above.  For the specific densities
where it is possible to cover the lattice with loops of a unique length, the
result is an ordered \textit{loop crystal}.  In some cases, however, a
covering by a unique length or by a combination of loops of two lengths, is
not possible.  We point out some such cases below.

\subsubsection{Checkerboard}

\begin{figure}[tb]
\centering{\includegraphics[width=0.9\columnwidth]{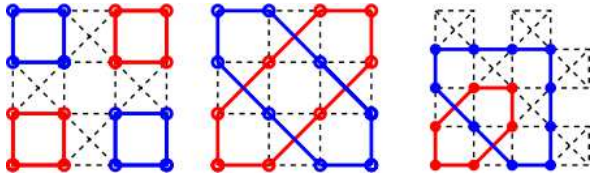}}
\caption{Schematic representation of the checkerboard lattice with 16 sites
and different loop coverings. 
\textit{Left:} 4 loops of length 4, corresponding to the ground state at
$\rho=1/4$ with 1 electron per loop. 
\textit{Center:} 2 loops of length 8, corresponding to the ground state at
$\rho=3/8$ with 3 electrons per loop. Between 1/4 and 3/8, a mixture of these
two configurations will occur, their respective ratio being set by the total
number of electrons $N_{e}=\rho N$. 
This arrangement of two loops of length 8 appears as a ``defect'' in a crystal of loops of length 4. 
\textit{Right:} a red loop of length 6 spanned by another blue loop; the
shortest way to close this blue loop requires 10 sites. 
It is thus impossible to have a mixture of loops of length
6 and 8 only.
}
\label{fig:loop48}
\end{figure}

We first focus on the regime $\rho \in [1/4;3/8]$ favoring loops of length 4
or 8 according to the Maxwell construction.  As illustrated in
figure~\ref{fig:loop48}, not only can the lattice be covered by these loops,
but the transformation from 4 loops of length 4 to 2 of length 8 is also
purely local and allows all possible ratios between these two phases in the
thermodynamic limit.  A pair of loops of length 8 (one made of up spins, the
other of down ones) cannot be separated if there are no other lengths than 4
and 8 in the system.  We shall call such pair a defect.  The defect concentration
is determined by $\rho$. Defects are not topological in the sense that they can
be created and annihilated locally, they can be placed anywhere in the background made of loops of length 4.

At quarter filling, there is one electron per loop of length 4, its energy is
$-2t$ and their total number is $N/4$. Each additional pair of electrons fills
the first excited level of the newly created defect (3 electrons per $\ell=8$
loop) and gives an energy $-4t\left(\sqrt{2}-1\right)$.
At a given $\rho$, the number of electrons added with respect to the loop
crystal at quarter filling is $(\rho N-N/4)=(4\rho-1)L^{2}$. The total energy
of the system between $\rho=1/4$ and $3/8$ is then
\begin{eqnarray}
E(\rho)=-2t\,L^{2}\,\left(1+(4\rho-1)(\sqrt{2}-1)\right)
\label{eq:nrj48}
\end{eqnarray}
Once normalized per number of sites $4L^{2}$, this expression is the dashed
line plotted on figure~\ref{fig:nrjrho}.  
We have verified this analytical result with finite-temperature
Monte Carlo simulations; some details of the method are in
Appendix \ref{appworm}. Extensive degeneracy is recovered in this region between 1/4 and 3/8.

At higher doping, above $\rho=3/8$, the Maxwell argument predicts a mixture of
loops of length 8 and 6.  However, as we can see in figure~\ref{fig:loop48}, a
single loop of length 6 imposes the presence of loops of length 10 at least.
This implies that this region of the phase diagram (obtained without
accounting for such lattice constraints) is further modified, in an as yet unknown manner.

\subsubsection{Pyrochlore}
\label{sec:pyrokag}

Analogous modification of the phase diagram of figure~\ref{fig:PD} is more severe for the pyrochlore, and
shows up already at $\rho=1/\ell_{min}=1/6$.  We have found that it is
impossible to cover the pyrochlore lattice with loops of length 6 only.
However, it is possible to do it for half of the system, as explained next.

The pyrochlore lattice can be seen as an alternative stack of kagome and triangular layers orthogonal to one of the global [111] axes. As depicted on figure~\ref{fig:trikag}(d), 2/3 of each  kagome layer can be filled with loops of length 6 (blue hexagons), while the other 1/3 of the kagome sites forms extensive winding loops along the [111]-axis, crossing alternatively the kagome and triangular layers. Since all blue sites form loops of length 6, putting one electron per loop provides a ground state configuration up to $\rho=1/12$ at least. This is a priori not the only one, but this proves its existence.

Indeed, despite an intensive search by complete enumeration of configurations respecting the ice-rules on the pyrochlore lattice, we have not detected any relevant loop crystal or mixture of them for systems up to 128 sites. A system of size 192 has been partially investigated, with the same outcome. The smallest occurrence of a single-length loop covering is
for 8 loops of length 16 in a system of 128 sites; however this is not a
relevant length according to our phase diagram of Figure \ref{fig:PD}. Thus,
the loop coverings on an actual pyrochlore are modified from
Figure \ref{fig:PD} in most or all of the density range $\rho\in[1/6,1/2]$.
The details of this modification remains an important open problem.


\section{Conducting channels or ``filaments''}  \label{sec:conduction}

In this section, we discuss the effect of the Coulomb phase and loop
structure on the conductivity of mobile electrons. Since the electrons are
confined to loops, they can conduct only if a loop connects one edge of
the sample to the opposite end.  Therefore, we study the number of such
sample-spanning loops, which, following Ref.~\onlinecite{Viret04a}, we refer
to as ``filaments''.

When there are no filaments (\textit{e.g.}  in loop crystals with only finite-length loops), the system is unambiguously an insulator. When there are filaments spanning the system, the system
cannot be immediately called a conductor, because the actual conductivity
will depend on scattering mechanisms exterior to our model.

In the first subsection below, we consider small dopings, where we present
Monte Carlo results for the average number of filaments as a function of
system size.  In the second subsection, we comment on the consequences of our
phase diagram.

\subsection{Filaments at low densities}

We first consider very low densities that the loop distribution can be assumed
to be largely unchanged from the $\rho=0$ case.  

In 2D, there is a small but constant number ($\approx{1.86}$) of  winding
loops on average in the thermodynamic limit~\cite{Jaubert11a}. There is thus
some probability to have filaments in the checkerboard case, but the number of
filaments does not grow with system size.

\begin{figure}
\centering\includegraphics[width=0.98\columnwidth]{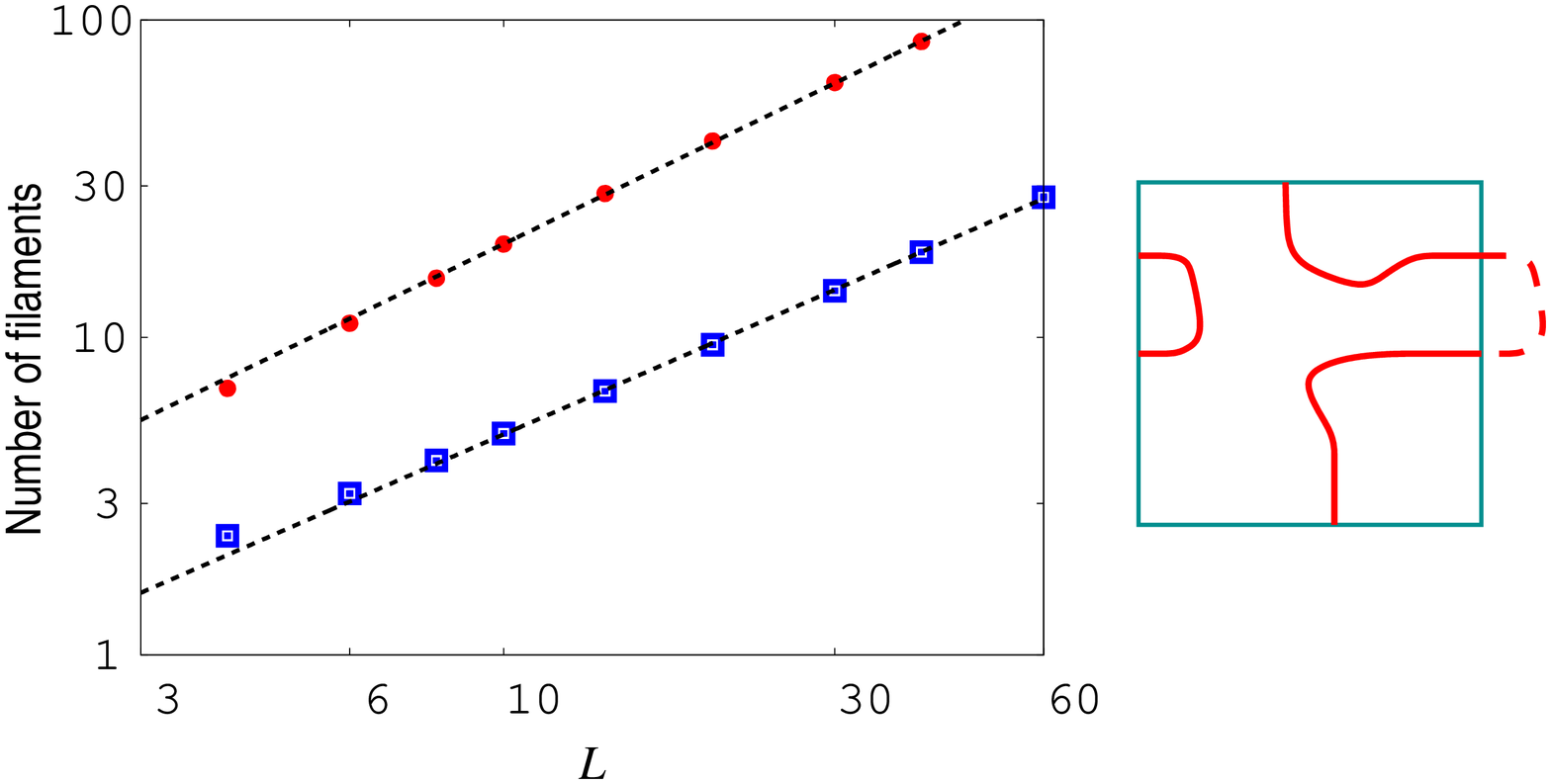}
\caption{
Average number of filaments, \textit{i.e.}  those segments of loops spanning the entire system from
one border to the opposite one, as a function of system size $L$.  We use two
definitions of filaments; they can either cross the orthogonal borders thanks
to the periodic boundary conditions $(\textcolor{red}{\bullet})$ or not
$(\textcolor{blue}{\blacksquare})$.  The cartoon on the right shows such a
boundary-crossing filament, which would be excluded in the
$\textcolor{blue}{\blacksquare}$ data.  Dashed lines are guides to the eye
for the linear behavior with $L$.  Both x- and y-axes are on a logarithmic scale.
}
\label{fig:avlNs}
\end{figure}

In 3D, the background of winding loops ensures that there are filaments whose
number grows with system size.  The data in Figure \ref{fig:avlNs} shows
that the number of filaments increases linearly with the linear size of the
cubic sample.

Since our simulations and loop counting are performed with periodic boundary
conditions, some of the filaments, while spanning the sample in one direction,
also cross one of the orthogonal boundaries.  (An example is shown in the
cartoon to the right of Figure \ref{fig:avlNs}.)  One can argue that this type
of filament would not contribute to conduction in a real-life cubic pyrochlore
sample.  Therefore, we show data both excluding and including this type of
filament, and they are seen to have the same power-law behavior.

Since the data including these filaments have better statistics, for finite $\rho$ we display the inclusive data, with the expectation that there is no qualitative difference.

\begin{figure*}[ht]
\centering\includegraphics[width=8cm]{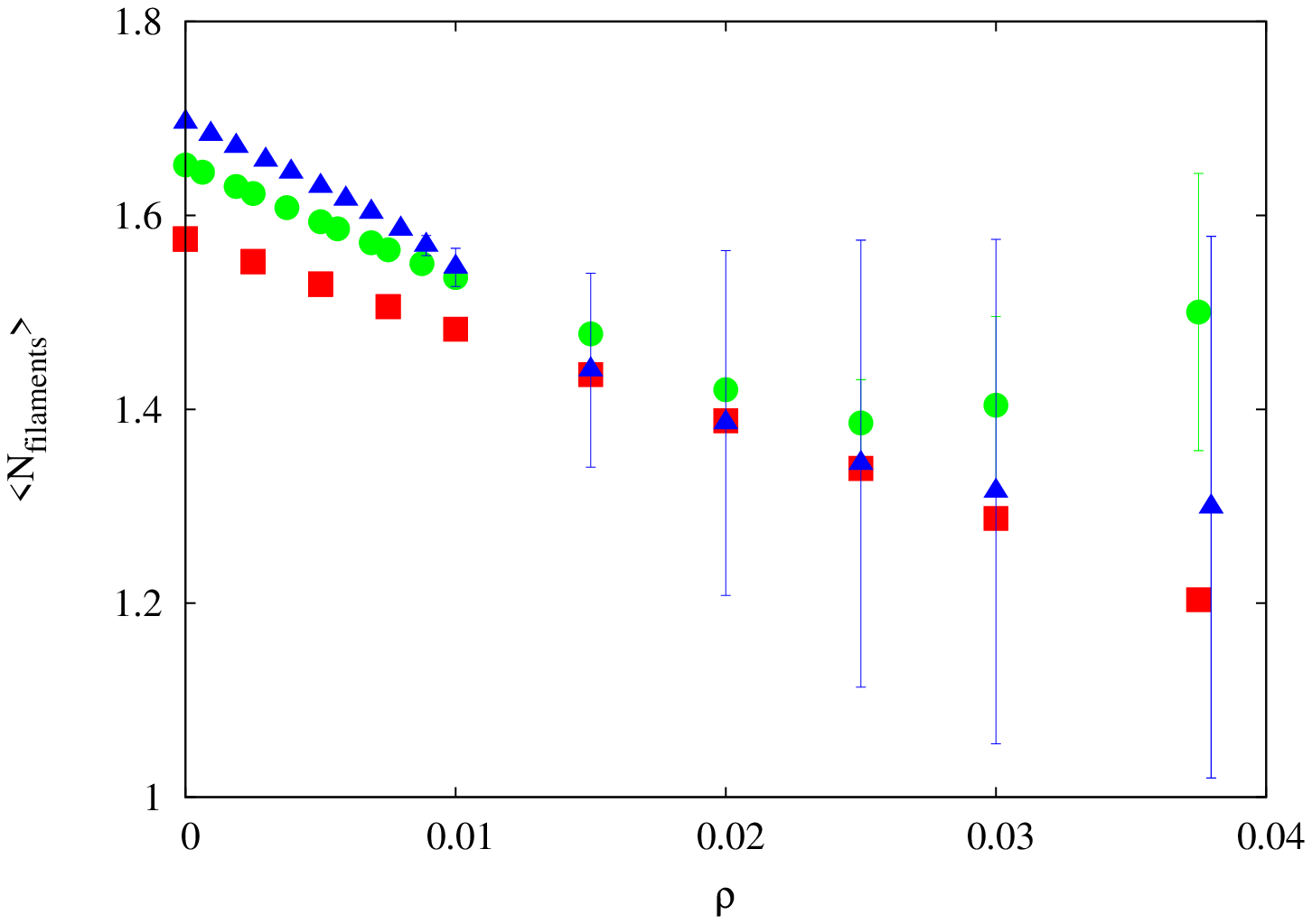}
\centering\includegraphics[width=8cm]{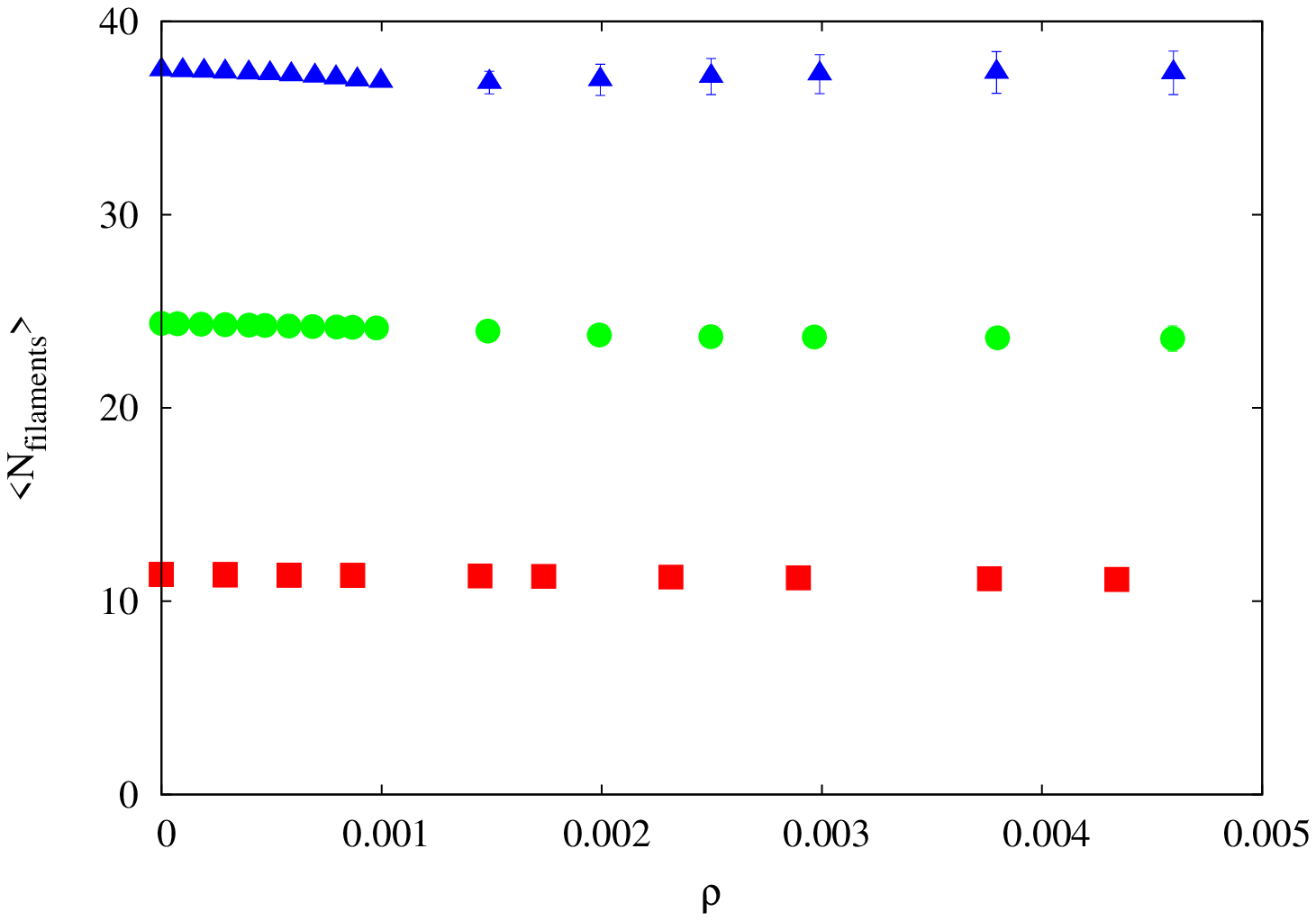}
\caption{Average number of filaments in 2D (\textit{left}) and 3D (\textit{right}), corresponding respectively to systems of sizes $L=10, 20, 40$ and $L=6, 12, 18$ (red squares, green dots, blue triangles) as a function of doping $\rho$.}
\label{fig:Nfil}
\end{figure*}

According to Fig.~\ref{fig:pdfelec3d}, itinerant electrons tend to make disappear extensive loops of length $L^{3}$ in 3D. Whether or not this prevents the formation of filaments is not as straightforward as it seems. Indeed, one could naively assume that the number of conducting channels will decrease and maybe even vanish in the thermodynamic limit. However the number of filaments remains approximately constant as plotted on figure~\ref{fig:Nfil}. This means that the average number of loops increases with doping via dividing the very long ones into smaller but nonetheless extensive loops spanning the system. In the low doping regime in 3D, the number of conducting channels is approximately \textit{independent} of $\rho$.

By contrast, in 2D, with relatively large error-bars, the number of filaments decreases but remains of $\mathcal{O}(1)$ as electrons are added.

\subsection{Conduction channels at larger fillings}

In Section \ref{sec_phasediagram}, we identified densities (for both 2D
and 3D) where the system is a loop crystal or is covered by loops of
two finite lengths only.  In such cases, there are no filaments, and the
system is truly insulating: lifting the frustration-induced degeneracy removes the non-locality of the loops. How the energetics (and resulting degeneracies) imposed by lattice constraints (see section~\ref{sec:lattice_constraints}) manifest themselves in transport properties is an intriguing separate question.

For some ranges of $\rho$, especially in 3D, the loop coverings indicated by
the Maxwell construction are disallowed by the lattice geometry
(section \ref{sec:lattice_constraints}).  In such cases it remains an open
question whether or not the lattice constraints result in loop coverings
including infinite (sample-spanning) loops.  Unfortunately, the issue of
conducting channels depends on the answer to this generally unresolved
question.

\section{Summary and outlook} 

We have analyzed the double exchange
model on the pyrochlore lattice in two and three dimensions. We have
chosen to consider a parameter range for which (i) magnetic
frustration is known to give rise to unconventional ground state
ensembles and (ii) where it is possible to make considerable
analytical progress by mapping the system onto an ensemble of loops,
the statistical properties of which are influenced by the addition of
the electrons.

We have identified a number of phenomena which depend on
features such as dimensionality, which determines whether or not there
exist loop segments winding around the system; or lattice structure,
which may frustrate the geometric packing of preferred loop lengths. 

The model studied here leaves unanswered a number of questions and
immediately suggests many generalizations and extensions. We have
worked in a limit of parameters such that the magnetic exchange
and anisotropy dominate over the Hund coupling which in turn dominates
the hopping integral. Our analysis applies to zero temperature.

It would be interesting to relax any of these choices, although
technically this may not be easy. In particular, given the presence of
gapless excitations on long loops, interesting low-temperature physics
may appear. Canting can give rise to non-trivial Berry phase physics,
and finite Hund's coupling will enable electrons to hop between loops.
For example, recent work at quarter filling on the pyrochlore Kondo lattice has shown the emergence of a chiral magnetic order in the weak-coupling regime~\cite{Chern10a}.

Finally, even in the parameter range discussed here, it will be
interesting to ask how different frustrated lattices shape up compared
to the pyrochlores. We devote the final paragraphs of this paper to
discussing this question for the case of triangle-based lattices, the
triangular Bravais lattice and the Archimedean kagome lattice.

\subsection{Other lattices: triangular and kagome} 

The Ising ground states observed for the pyrochlore lattice have
vanishing total spin on each tetrahedron. These `ice rules' ensure the
existence of the Coulomb phase and states obeying them amount to a
moderate zero-point entropy of less than a third of that of a free
spin. By contrast, the zero-point entropy of the triangular Ising
magnet is not far from half of that of a free spin, while that of the
kagome magnet is over 70\% of $\log2$.

Most fundamentally, the single triangle is relatively much more
degenerate than a tetrahedron, with 6 out of 8 (rather than out of 16)
states being ground states. The triangle states have varying
magnetizations of $\pm 1$ (whereas states obeying the ice rule have a
unique magnetization, unless one tunes a field to a transition between
magnetization plateaux~\cite{Moessner09a}).

To move towards the full triangle-based lattices, it is worth noting
that for any Ising antiferromagnet, the `hopping network' formed by
neighbouring aligned spins can have a coordination of at most half
that of $z$, the coordination of the underlying lattice -- otherwise
it would be energetically favorable to flip the highly coordinated
spin. The concept of the hopping network generalizes the loops on
which the electrons hop on the planar and three dimensional pyrochlores. 

It now turns out that triangle and kagome lattices behave entirely
differently from pyrochlore in both two and three dimensions, on
account of the nature of their frustrated ground state ensembles.

\subsubsection{Triangular lattice} 

The ground-state degeneracy of the triangular lattice is immediately
lifted by the addition of even a single hole. This result is entirely
analogous to the frustrated Nagaoka theorem presented in
Ref.~\onlinecite{Moessner00a}, in the context of the magnetic
supersolid discussed there, and it is also connected to the triangular
Bosonic supersolids~\cite{Murthy97a,Heidarian05a,Wessel05a,Melko05a}. 

\begin{figure}[ht]
\centering\includegraphics[width=8cm]{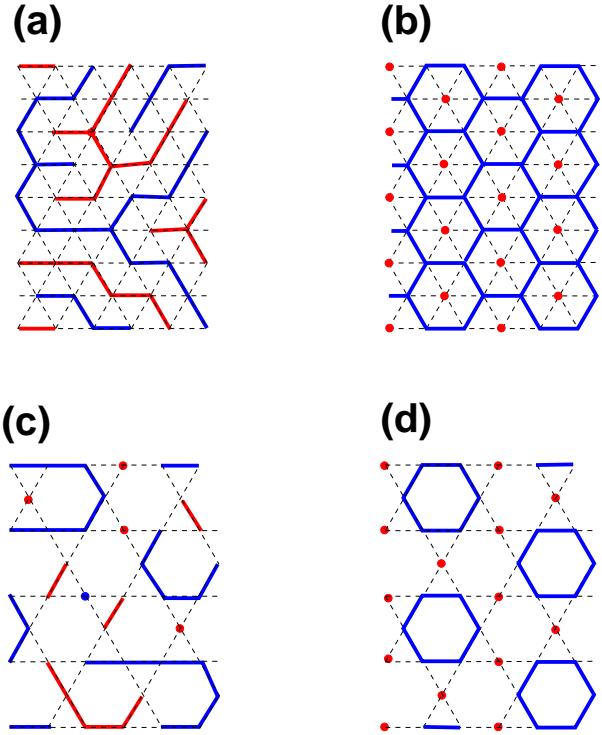}
\caption{Configurations on the triangular (top) and kagome (bottom) lattices respecting the antiferromagnetic frustrated constraints. \textit{(a)} and \textit{(c)} are random configurations, while \textit{(b)} is the conducting hopping network minimizing the kinetic energy on the triangular lattice, and \textit{(d)} is the insulating state maximizing the number of closed loops on kagome. Panel \textit{(d)} also illustrates the arrangement of loops described in section~\ref{sec:pyrokag} in the  two dimensional kagome layers which are part of the three dimensional pyrochlore lattice.
}
\label{fig:trikag}
\end{figure}

As $z=6$ for the triangular lattice, the hopping network of aligned
spins no longer has coordination two as in the loops of the pyrochlore
lattice. In fact, the coordination of a site of the hopping network no
longer even needs to be uniform, so that there may be dangling or even
isolated sites, as on Fig.~\ref{fig:trikag}.(a). 
The coordination can range all the way from zero (for
a spin surrounded by a hexagon of oppositely aligned spins) to
maximally $z/2=3$.

The latter happens when the hole sits on a site experiencing zero net
exchange field from its six neighbors. There exists a unique state
(pictured in Fig.~\ref{fig:trikag}.(b)) in which there is a network of three-fold
coordinated sites. This state breaks translational symmetry by
tripling the unit cell as well as time-reversal symmetry as it
corresponds to a state with a magnetization of a third of the
saturated value. We have not studied what happens to a finite doping
but a Fermi liquid regime on the hexagonal backbone at
low doping looks likely. This would imply a conducting state,
with the possibility of additional low-energy excitations in the form
of defects of the hexagonal backbone.

\subsubsection{Kagome lattice} 

The situation on the kagome lattice is different still. Its magnetic ground
state ensemble is very short-range correlated, unlike that of the triangular
lattice, which has algebraic correlations. With $z=4$ for the kagome lattice,
the hopping network can no longer branch but it can now have dangling links or isolated spins (coordination 1 and 0, respectively, as illustrated on Fig.~\ref{fig:trikag}.(c)).

The hopping networks thus consist of loop segments which need no
longer close on themselves, minimally containing only one spin but not
bounded above in the thermodynamic limit. The length distribution is,
however, unlikely to contain long loops as this would only happen if
each the magnetization of all the triangles along the loop segment
has the same sign, at considerable cost in entropy. 

Energetically, it is of course again most advantageous to have closed
loops, as electrons on them gain hopping energy $-2|t|$. On the kagome
lattice, such loops are readily constructed. The shortest ones are
obtained by arranging spins to be aligned around a hexagon. Their
densest packing is obtained for a state which breaks translational
symmetry, tripling the unit cell and incorporating a 1/3 magnetization
as was the case of the triangular lattice above, see Fig.~\ref{fig:trikag}.(d). However,
this state -- which is the unique ground state at electron density
$\rho=1/9$ is now an insulating one -- all hopping paths for the
electrons close back on themselves after six steps.

\begin{acknowledgements}

We thank J.~Chalker, C.~Henley, and M.~Viret for useful
discussions.  

\end{acknowledgements}

\appendix

\section{Worm algorithm}
\label{appworm}

\paragraph{Worm algorithm without electrons.}

\begin{figure}[thbp]
\centering\includegraphics[width=0.8\columnwidth]{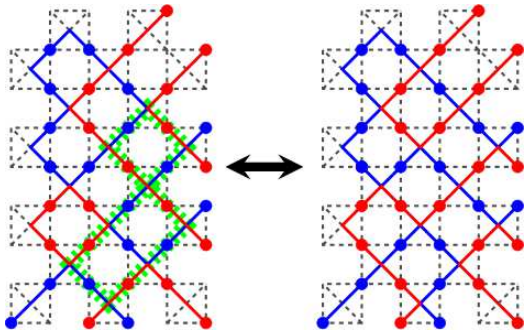}
\caption{
On the left, the same loop configuration as in figure~\ref{fig:latt}. The
thick dashed green line represents a possible worm, as built by our algorithm;
flipping the spins along this worm does not break the ice-rules (see on the
right) and allows us to visit the configurational space of the Coulomb phase,
ensuring ergodicity.
}
\label{fig:latt_worm}
\end{figure}

In our definition, a loop is made of nearest neighbour spins pointing in the
same direction: it is uniquely defined and possesses an up or down flavor. On
the other hand, a worm consists of alternating up/down/up/down/... spins;
through each vertex of the premedial lattice (square in 2D or diamond in 3D);
see figure~\ref{fig:latt_worm}.  The worm can randomly choose between two paths
energetically equivalent and eventually hit its initial position; the worm is
then closed.  Reversing all spins in the worm gives way to a new configuration
in the Coulomb phase (figure~\ref{fig:latt_worm}). This method ensures both
ergodicity and detailed balance to the algorithm in absence of
electrons~\cite{Barkema98a}.

Worms, as defined above, are not self-avoiding, \textit{i.e.}  they can erase their own
path. If a worm goes an odd number of times through the same site, the
corresponding spin will be flipped during the Monte Carlo update; if it
happens an even number of times, the spin shall not be flipped.

\paragraph{Decorrelation.}

Since we are dealing with Ising spins, two random configurations roughly
differ by one half of the system sites on average. Hence two configurations
separated by $k$ worm updates can be considered decorrelated if close to one
half of the system sites are different: we arbitrarily chose 45\%. Previous
work~\cite{Jaubert11a} has shown that a finite constant number of worm updates
($\sim 10$) was enough to decorrelate the system in 3D, but an increasing
number with system size $L$ was necessary in 2D. This is related to P\'olya's
theorem stating that a random walk (similar to the worm here) in 3D is
transient (finite probability never to come back to the origin), while it is
recurrent in 2D (it always comes back)~\cite{Hughes95a}.

\paragraph{Worm algorithm with electrons -- entropic regime.}

In the entropic regime, \textit{i.e.}  at low doping, there is at most one electron per
loop. A correct sampling can then be done by using the above worm algorithm
free of electrons, and rejecting all configurations with more electrons
$N_{e}$ than loops $N_{\ell}$. However, the distribution of number of loops
per configuration being Gaussian (see figure~\ref{fig:entropy2}), the density
of electrons acts as a cutoff and above a certain threshold of the order of
$1/\langle \ell \rangle$, almost all configurations are rejected and it
becomes impossible to get good statistics.

In order to take into account the influence of itinerant electrons in the
entropy, we chose to weigh any loop configuration by the number of possible
combinations to distribute $N_{e}$ electrons in $N_{\ell}$ loops, namely the
binomial coefficient $C_{N_{\ell}}^{N_{e}}$.

\paragraph{Worm algorithm with electrons -- energetic regime.}

In the energy regime, \textit{i.e.}  at intermediate or high doping, there is more than
one electron per loop. The above method based on the worm algorithm without
electrons become inefficient and finite temperature Monte Carlo simulations
are necessary.

Starting from a given configuration, a worm update is proposed.  The loop
histogram of the states before and after the proposed update are computed, and
the corresponding electron energies are calculated after filling the energy
levels with $N_{e}$ electrons.  The worm update is then accepted or not via a
Metropolis argument.

\end{document}